\pdfoutput=1
\documentclass[nofootinbib,aps]{revtex4}
\usepackage{amssymb,amsmath,graphicx,color,microtype}
\usepackage{enumerate}
\usepackage{slashed}
\usepackage{hyperref}
\hypersetup{
	colorlinks=true,
	linkcolor=blue,
	filecolor=blue,      
	urlcolor=blue,
	pdftitle={Sharelatex Example},
	bookmarks=true,
	citecolor=blue
}

\usepackage[paperwidth=210mm,paperheight=297mm,centering,hmargin=2cm,vmargin=2.5cm]{geometry}

\begin{document}
	\title{Warm Quasi-Single Field Inflation} 
	\author{Xi Tong$^1$}
	\email{tx123@mail.ustc.edu.cn}
	\author{Yi Wang$^{2,3}$}
	\email{phyw@ust.hk}
	\author{Siyi Zhou$^{2,3}$}
	\email{szhouah@connect.ust.hk}
	\affiliation{${}^1$School of Physics, University of Science and Technology of China, \\
		Hefei, Anhui 230026, China}
	\affiliation{${}^2$Department of Physics, The Hong Kong University of Science and Technology, \\
		Clear Water Bay, Kowloon, Hong Kong, P.R.China}
	\affiliation{${}^3$Jockey Club Institute for Advanced Study, The Hong Kong University of Science and Technology, \\
		Clear Water Bay, Kowloon, Hong Kong, P.R.China}
	
	\begin{abstract}
		We study quasi-single field inflation with a warm inflation background. The thermal effects at small scales can sufficiently enhance the magnitude of the primordial standard clock signal. This scenario offers us the possibility of probing the UV physics of the very early universe without the exponentially small Boltzmann factor when the mass of the isocurvaton is much heavier than Hubble. The thermal effects at small scales can be studied using the flat space thermal field theory, connected to an effective description using non-Bunch-Davies vacuum at large scales, with large clock signal.
	\end{abstract}
	
	\maketitle
	
	\section{Introduction}
	\numberwithin{equation}{section}
	The characteristic features of massive fields with masses $m\sim H$ during inflation (quasi-single field inflation and related scenarios) attracted much interest recently \cite{Chen:2009we,Chen:2009zp,Tolley:2009fg,Achucarro:2010jv,Jackson:2010cw,Cremonini:2010ua,Baumann:2011su,Baumann:2011nk,Achucarro:2012sm,Assassi:2012zq,Sefusatti:2012ye,Norena:2012yi,Chen:2012ge,Pi:2012gf,Achucarro:2012yr,Gwyn:2012mw,Noumi:2012vr,Cespedes:2013rda,Gong:2013sma,Emami:2013lma,Castillo:2013sfa,Kehagias:2015jha,Liu:2015tza,Achucarro:2015rfa,Arkani-Hamed:2015bza,Dimastrogiovanni:2015pla,Maldacena:2015bha,Schmidt:2015xka,Chen:2015lza,Bonga:2015urq,Chen:2016cbe,Chen:2016nrs,Flauger:2016idt,Lee:2016vti,Chen:2016qce,Liu:2016aaf,Delacretaz:2016nhw,Meerburg:2016zdz,Chen:2016uwp,Chen:2016hrz,Jiang:2017nou,Chen:2017ryl,An:2017hlx,Tong:2017iat,Iyer:2017qzw,An:2017rwo,Kumar:2017ecc,Riquelme:2017bxt}. Fields with $m>H$ can arise from string theory, for example the string oscillatory modes or Kluza-Klein modes. And fields with $m\sim H$ can be generated from Standard Model uplifting \cite{Chen:2016nrs,Chen:2016uwp,Chen:2016hrz}, supersymmetry breaking \cite{Baumann:2011nk,Delacretaz:2016nhw} and non-minimal coupling. Apart from that, those massive fields are also standard clocks to model-independently distinguish different very early universe scenarios \cite{Chen:2015lza}. If the signature of these modes can be observed, it provides hints about particle physics, string theory and the evolution history of the universe. 
	
	The massive fields with mass $m$ and spin $s$ can produce an oscillatory signature on the squeezed limit non-Gaussianities as $(k_3/ k_1)^{3/2+i \mu} P_s (\cos \theta)$ \cite{Chen:2009we,Chen:2009zp,Baumann:2011nk}, where $\mu=\sqrt{m^2/H^2-9/4}$. This squeezed limit behavior helps us separating the signal from the massive fields from the conventional single field backgrounds such as slow roll inflation with $(k_3/ k_1)^0$ and a modified sound speed with $(k_3/k_1)^2$. However, the conventional quasi-single field inflation relies on the production of massive particles by de Sitter spacetime through a Bogolyubov transformation. Equivalently, massive particles are created by the Hubble temperature heat bath, thus are Boltzmann suppressed by the Hubble scale $\exp(-2\pi \mu)$. The actual clock signal are suppressed by $\exp(-\pi \mu)$. Because the clock signal comes from the  interference of the positive frequency part and the negative frequency part. One can regard this as the cosmological double slit experiment \cite{Arkani-Hamed:2015bza}. And due to the Boltzmann suppression, fields with large masses are almost impossible to leave a detectable signature on the non-Gaussianities. Therefore, it is difficult to probe physics at extremely high energies though the conventional way. Yet things are different in the warm inflation scenario.
	
	Warm inflation \cite{Berera:1995wh,Berera:1995ie} was initially proposed to generate radiation without reheating. During inflation, the universe has a finite temperature $T\gg T_{dS}=H/(2\pi)$. The zero mode of the inflaton can decay into radiation continuously. The small scale physics is governed by flat spacetime finite temperature field theory since the curvature of spacetime is negligible. Upon this set up the decay rate of the zero mode inflaton and the production rate of radiation can be calculated. The particle production process backreacts on the inflaton background and a large dissipation can be generated and extra friction is added to the rolling inflaton.
	
	The small scale UV region is like a furnace powered by the rolling inflaton. A heat bath of particles come out of it and are diluted by the inflationary expansion. The expansion also causes the stretching of their wavelength and thus the decrease in energy. As these particles approach the Hubble scale, they begin to feel another temperature which is rooted deep within the structure of de Sitter spacetime. As soon as the mode become low-energy states, one can say that it essentially exit the thermal equilibrium in the UV. Instead, it enters the domain of conventional quasi-single field inflation governed by Hubble temperature. Therefore its evolution should be described by the conventional mode functions. 
	
	The picture seems operational except for one potential loophole. At first sight, we seem to be using a thermal gas of massive particles to do a double slit experiment. It looks like impossible to obtain any interference pattern on the screen since a bunch of thermalized particles could not be coherent. Statistical average soon flattens the patterns. But a closer look reveals a remedy. Out of a furnace in the UV there could be many particles. Those that does not come from the same process must have irrelevant phases and will be averaged out in the end. However, particles produced during the exit from thermal equilibrium essentially interfere with themselves in a similar way that they do in cold quasi-single field inflation. In other words, the double slit is no longer the Hubble horizon but the thermal horizon of the UV furnace.
	
	The physics in the IR region is analogous to what happens in the conventional cold quasi-single field inflation. The clock signal is generated via the resonance mechanism of the massive field and the inflaton. This mixing can happen in the situation such as, the inflaton has a constant turning trajectory. 
	In this paper, we consider the following two operators in the Lagrangian
	\begin{align} 
	L_2 =  a^3 C_2 \int d^3 x \sigma \phi', \quad L_3 =  a^2 C_3 \int d^3 x  \phi'^2\sigma~,
	\end{align}
	both of which originates from a dimension five operator, 
	\begin{align}\label{EFTsetup}
	\mathcal O_5 = -\frac{1}{2\Lambda_0} (\partial \phi)^2 \sigma~.
	\end{align}
	As a result, the power spectrum and the bispectrum of our model can be generated via this effective field theory operator.
	
	This paper is organized as follows: In Section~\ref{formalsection}, by considering the mass shift when an individual comoving mode exits the thermal horizon, we formally construct a non-BD massive mode to mimic the thermal effects in the UV furnace. In Section~\ref{QMsection}, we draw an analogy of the particle production rate to the transmission rate in a quantum mechanical system and calculate the probability of particle production. The result gives an intuitive understanding of the particle production rate from the thermal bath. We then compare a smooth exit and an abrupt exit. We found that the particle production rate is consistent using these two methods. In Section~\ref{3ptsection} and \ref{4ptsection}, we compute the squeezed limit of bispectrum and the collapsed limit of trispectrum. Clock signals without mass Boltzmann suppression are found. 
	
	\section{A formal construction}\label{formalsection}
	Traditionally we use the in-in formalism \cite{Weinberg:2005vy} (see also recent reviews \cite{Chen:2010xka,Wang:2013eqj}) to calculate the late time expectation value of an operator $\mathcal O$ in the state of our universe $\Omega$. 
	\begin{equation}\label{ininobservable}
	\langle\mathcal{O}\rangle=\langle\Omega|\bar{T}e^{i\int_{-\infty}^{0}d\tau' H_I( \tau')}\mathcal{O}(0)Te^{-i\int_{-\infty}^{0}d\tau H_I(\tau)}|\Omega \rangle~.
	\end{equation}
	where $\bar T$ and $T$ denotes anti-time ordering and time ordering, respectively. $H_I$ is the interacting Hamiltonian. $\tau$ is the conformal time.
	
	The integral in the evolution operator extends infinitely deep into the UV regime. But now the UV physics becomes complicated due to the existence of thermal radiation. Our solution is to use finite temperature field theory in the UV above the thermal horizon and use in-in formalism (Schwinger-Keldysh formalism) in the IR. Then we sew them together at the thermal horizon. 
	
	\begin{figure}[htbp] 
		\centering 
		\includegraphics[width=8cm]{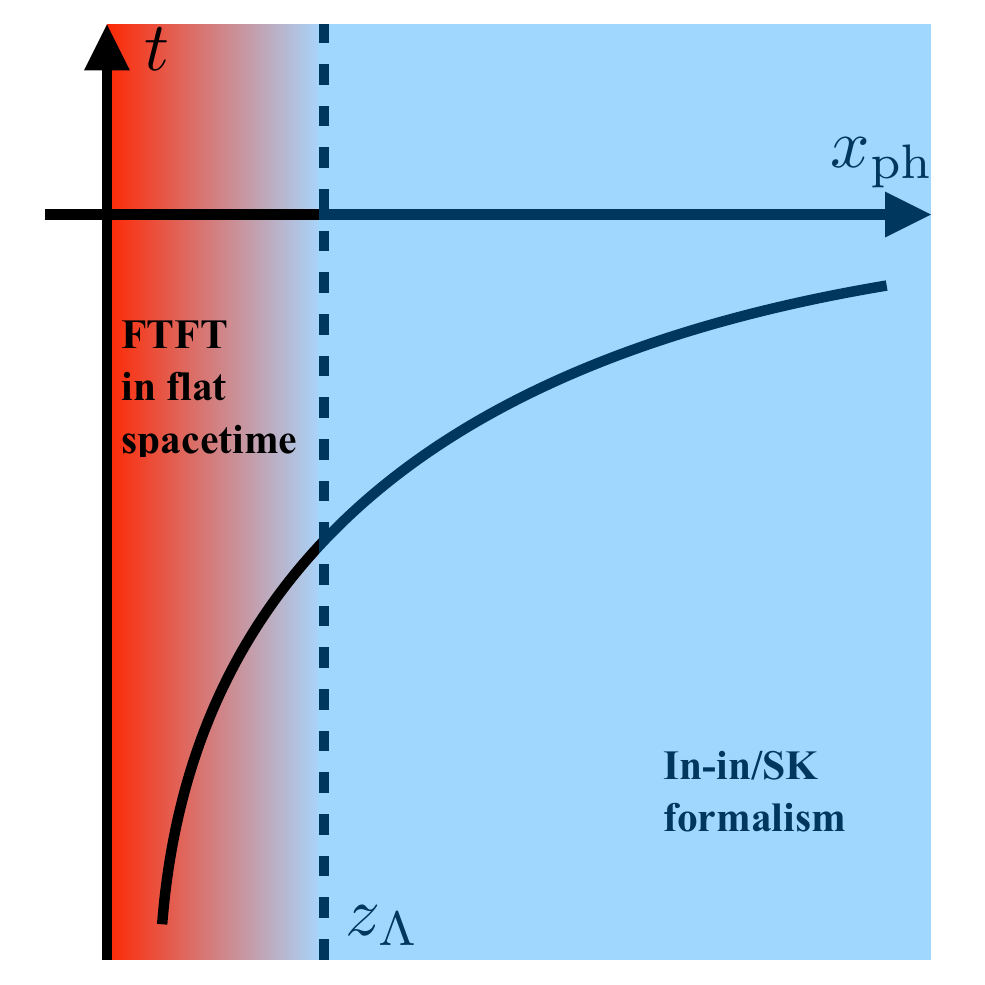}   
		\caption{We use different formalisms at different scales. The axis are physical time and distance. The curve stands for the exponential growth of the physical wavelength of a given comoving mode. Inside $z_\Lambda$, the mode behaves like a particle in flat spacetime thermal equilibrium. Outside it behaves like a quantum wave in curved spacetime.} 
	\end{figure}
	
	We use $|z_\Lambda\rangle$ to denote the state of the universe for the given comoving modes at that scale\footnote{The subscript $\Lambda$ here should not be confused with the EFT scale $\Lambda_0$ in the operator $\mathcal{O}_5$ in (\ref{EFTsetup}).}. How the states evolve above $z_\Lambda$ is in the control of thermal field theory in nearly flat spacetime. In the interaction picture, an observable $\mathcal{O}$ has a quantum expectation
	\begin{equation}
	\langle\mathcal{O}\rangle=\langle z_\Lambda|\bar{T}e^{i\int_{\tau_\Lambda}^{0}d\tau' H_I(\tau')}\mathcal{O}(0)Te^{-i\int_{\tau_\Lambda}^{0}d\tau H_I(\tau)}|z_\Lambda \rangle~.
	\end{equation}
	Now we ask if the problem can be solved in another way. We modify the interaction Hamiltonian and add extra effective vertices into it for $z>z_\Lambda$. This time the evolution begins at $-\infty$ and we assume a Bunch-Davies vacuum initial state. We call our new Hamiltonian $\tilde{H}_I(\tau)$.
	\begin{equation}
	\langle\mathcal{O}\rangle=\langle\Omega|\bar{T}e^{i\int_{-\infty}^{0}d\tau' \tilde{H}_I(\tau')}\mathcal{O}(0)Te^{-i\int_{-\infty}^{0}d\tau \tilde{H}_I(\tau)}|\Omega \rangle~.
	\end{equation}
	The requirement that this two methods be equivalent gives
	\begin{equation}
	Te^{-i\int_{\tau_\Lambda}^{0}d\tau H_I(\tau)}|z_\Lambda\rangle=Te^{-i\int_{-\infty}^{0}d\tau \tilde{H}_I(\tau)}|\Omega \rangle~,
	\end{equation}
	or simply
	\begin{eqnarray}
	\nonumber|z_\Lambda\rangle=S|\Omega \rangle~.
	\end{eqnarray}
	The operator $S$ is defined as
	\begin{equation}
	S\equiv\left(\bar{T}e^{i\int_{\tau_\Lambda}^{0}d\tau H_I(\tau)}\right)\left(Te^{-i\int_{-\infty}^{0}d\tau \tilde{H}_I(\tau)}\right)~.
	\end{equation}
	Now we split the integral interval and use the Baker-Campbell-Hausdorff formula,
	\begin{equation}
	e^{A+B}=e^A e^B e^{-\frac{1}{2}[A,B]}e^{\frac{1}{6}\left(2[B,[A,B]+[A,[A,B]]]\right)}\times\cdots~.
	\end{equation}
	We get
	\begin{eqnarray}
	\nonumber S&=&\left(\bar{T}e^{i\int_{\tau_\Lambda}^{0}d\tau H_I(\tau)}\right)\left(Te^{-i\int_{\tau_\Lambda}^{0}d\tau \tilde{H}_I(\tau)-i\int_{-\infty}^{\tau_\Lambda}d\tau \tilde{H}_I(\tau)}\right)\\
	&=&\left(\bar{T}e^{i\int_{\tau_\Lambda}^{0}d\tau H_I(\tau)}\right)\left(Te^{-i\int_{\tau_\Lambda}^{0}d\tau H_I(\tau)-i\int_{-\infty}^{\tau_\Lambda}d\tau \tilde{H}_I(\tau)}\right)\label{assume}\\
	\nonumber&=&\left(\bar{T}e^{i\int_{\tau_\Lambda}^{0}d\tau H_I(\tau)}\right)\left(Te^{-i\int_{\tau_\Lambda}^{0}d\tau H_I(\tau)}\right)\left(Te^{-i\int_{-\infty}^{\tau_\Lambda}d\tau \tilde{H}_I(\tau)}\right)\left(e^{\frac{1}{2}[\int_{\tau_\Lambda}^{0}d\tau H_I(\tau),\int_{-\infty}^{\tau_\Lambda}d\tau^\prime \tilde{H}_I(\tau^\prime)]}\right)\times\cdots\\
	&=&\left(Te^{-i\int_{-\infty}^{\tau_\Lambda}d\tau \tilde{H}_I(\tau)}\right)\times\left( 1+\mathcal{O}(\lambda^2,C_2^2,C_3^2,\dots)\right)~.\label{Sfinal}
	\end{eqnarray}
	In (\ref{assume}) we used the assumption that the effective interactions disappear after the modes exit the thermal horizon. So
	\begin{equation}
	\tilde{H}_I(\tau)=H_I(\tau)~~~\text{for}~~~\tau>\tau_\Lambda~.
	\end{equation}
	The high-order terms in (\ref{Sfinal}) are of order $\mathcal{O}\left(\lambda^2,C_2^2,C_3^2,\dots\right)$ while the terms that we care about in the evolution operator is of the first order. So for simplicity, we omit the higher terms and write
	\begin{equation}
	S\approx T\exp\left(-i\int_{-\infty}^{\tau_\Lambda}d\tau \tilde{H}_I(\tau)\right)~.
	\end{equation}
	
	The next thing to ponder is what new terms there are in the interaction Hamiltonian $\tilde{H}_I$. Since this part of interaction history lies deep in the UV and in a thermal bath with a high temperature $T\gg H$. Since the energy scale we are considering is much higher than the curvature of spacetime, the universe appears to be static for the physical processes, and it is appropriate to use flat spacetime finite temperature field theory. In a finite temperature field theory, the propagator of a field is modified by the thermal effects. In the real-time formalism, a free scalar particle propagates as
	\begin{equation}
	D_F^\beta(k)=\frac{i}{-k^2-m^2+i\epsilon}+\frac{2\pi}{e^{\beta E_k}-1}\delta\left(-k^2-m^2\right)~,
	\end{equation}
	where $\beta=1/T$. By calculating the self-energies of a species of particle, one can obtain the mass renormalization and the decay rate to other species of particles. As an example. for a toy model $\sigma^4$ theory, the mass correction at one loop order is easily worked out to be
	\begin{eqnarray}
	\Delta m^2(T)=\frac{\lambda T^2}{24}\times\left(1+\mathcal{O}(\frac{m}{T})\right)~.
	\end{eqnarray}
	We consider two other couplings that contributes to the thermal mass and list them below in Table.~\ref{Masstable}. The detailed computation is in Appendix~\ref{AppendixA}.
	\begin{figure}[htbp] 
		\centering 
		\includegraphics[width=13cm]{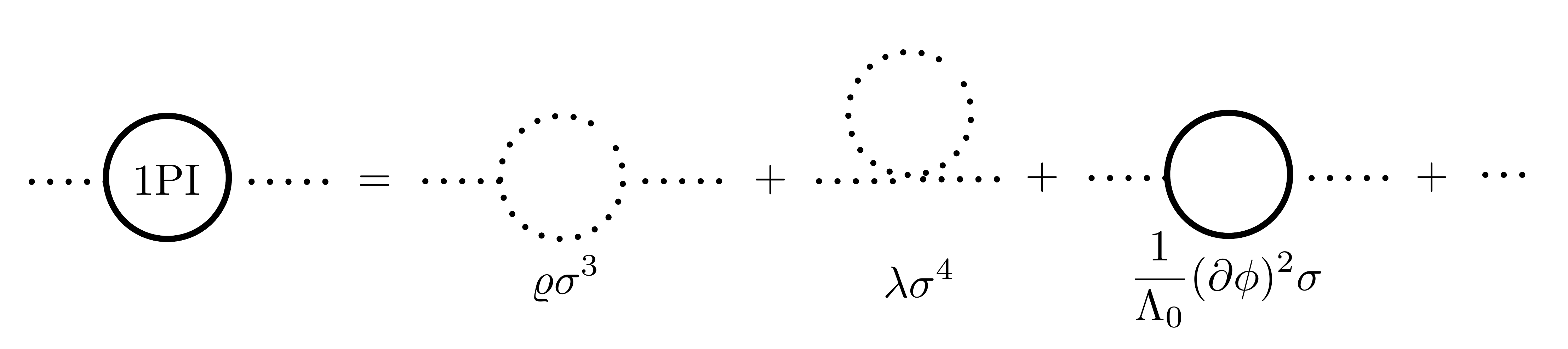}   
		\caption{$\sigma$ self-energy at one loop order.} 
		\label{fig:Pzeta}
	\end{figure}
	\renewcommand\arraystretch{1.5}
	\begin{table}[h!]
		\centering
		\begin{tabular}{ c| c c c}
			\hline\hline
			Interaction ~~~~~~&~~~~~~ $\frac{\varrho}{3!} \sigma^3$ ~~~~~~&~~~~~~ $\frac{\lambda}{4!} \sigma^4$   ~~~~~~&~~~~~~ $\frac{1}{2\Lambda_0} (\partial_\mu\phi)^2 \sigma$   ~~~~~~\\  
			\hline
			Mass correction $\Delta m^2(T)$ ~~~~~~&~~~~~~ $-\frac{\varrho^2 T}{30\pi m} $ ~~~~~~&~~~~~~ $\frac{\lambda T^2}{24}$  ~~~~~~&~~~~~~ $-\frac{m^2 T^2}{48 \Lambda_0^2}$ ~~~~~~\\
			\hline\hline
		\end{tabular}\caption{Thermal corrections to the mass due to different interactions.}\label{Masstable}
	\end{table}
	
	There are other corrections to the coupling constants but this is the one we mainly focus on as we will see in a moment. As a simplest example, the mass shift $\Delta m^2(T)$ is a constant, that is, if there is a sudden exit from thermal equilibrium at scale $z_\Lambda$, the mass of the $\sigma$ field changes suddenly. We call this the abrupt exit case. A simple and intuitive understanding of the abrupt case will be provided in Section.~\ref{QMsection}. In reality, the exit from thermal equilibrium at large scale happens smoothly and gradually. Therefore $\Delta m^2(T,z)$ must be a function of time for each individual mode. We set up a parametrization of $\Delta m^2(T,z)$ to address this smooth exit case also in Section.~\ref{QMsection}. For now, let us regard $\Delta m^2(T)$ as a constant to simplify the discussions.
	
	Another point that needs to be mentioned is that the nearly massless inflaton is protected by an approximate shift symmetry and we expect a negligible correction to its mass\footnote{Actually due to extra dissipation in warm inflation background, a not-so-flat inflaton potential is also acceptable. Nevertheless, the scenario involving inflaton masses only increases non-BD effects and make observables more observable. For the sake of demonstration, we focus on the simplest scenario here.}. So the most significant term in $\tilde{H}_I$ is the mass correction to the massive field,
	\begin{equation}\label{interactionhamiltonian}
	\tilde{H}_I=\int d^3x\left(\frac{1}{2}\Delta m^2(T)a^4\sigma^2+\text{usual stuff}\right)\approx\frac{1}{2}\int d^3x \Delta m^2(T) a^4\sigma^2~.
	\end{equation}
	Here we dropped the other interactions because we assume the massive field mass correction to be significantly larger than the other interactions. Yet it is still small enough to solve perturbatively. The $S$ operator can now be expressed as
	\begin{equation}\label{Sapprox}
	S=T\exp\left(-\frac{i}{2}\int_{-\infty}^{\tau_\Lambda} d\tau d^3x \Delta m^2(T)a^4\sigma^2\right)~.
	\end{equation}
	We are working in the interaction picture, hence a mode expansion can be used:
	\begin{equation}
	\sigma_\mathbf{k}(\tau)=v_k(\tau)a_\mathbf{k}+v_k(\tau)^*a_{-\mathbf{k}}^\dagger~.
	\end{equation}
	Insertion into (\ref{Sapprox}) yields
	\begin{eqnarray}
	S&=&T\exp\left(-\frac{1}{2}\Delta m^2(T)i\int_{-\infty}^{\tau_\Lambda} d\tau \frac{d^3k}{(2\pi)^3} a^4\left(v_k a_\mathbf{k}+v_k^*a_{-\mathbf{k}}^\dagger\right)\left(v_k a_\mathbf{-k}+v_k^*a_{\mathbf{k}}^\dagger\right)\right)~.
	\end{eqnarray}
	In inflation, the scale factor is given by $a(\tau)=-1/H\tau$. And the mode function follows from the equation of motion of the free massive fields,
	\begin{equation}\label{modefun}
	v_k=-i\frac{\sqrt{\pi}}{2}e^{i\pi\left(\frac{i\mu}{2}+\frac{1}{4}\right)}H(-\tau)^{3/2}H^{(1)}_{i\mu}(-k\tau),\quad \mu = \sqrt{\frac{m^2}{H^2}-\frac{9}{4}}~.
	\end{equation}
	In the extreme UV region $m\ll H z_\Lambda<H z$, the massive fields are relativistic and the mode function degenerates into the usual BD solution for inflatons,
	\begin{equation}
	v_k\approx \frac{H}{\sqrt{2k^3}}(1+ik\tau)e^{-ik\tau}\approx -\frac{iHz}{\sqrt{2k^3}}e^{iz}~.
	\end{equation}
	The mode function squared $v_k v_k$ contains a factor of $k^{-3}$, and will be canceled by the change of variable in the integration. This is crucial to ensure scale invariance in the final $n$-point functions. After performing the time integral, the $S$ operator becomes
	\begin{eqnarray}
	\nonumber S&=&\exp\left\{\frac{1}{2}\int\frac{d^3k}{(2\pi)^3} \left[\varpi a_\mathbf{k}a_\mathbf{-k}-\varpi^{*}a_{\mathbf{k}}^\dagger a_{-\mathbf{k}}^\dagger-i|\xi|\left(a_\mathbf{k}a_\mathbf{k}^\dagger+a_\mathbf{k}^\dagger a_\mathbf{k}\right)\right]\right\}~.
	\end{eqnarray}
	Here we used abbreviation
	\begin{eqnarray}\label{varpi}
	\varpi&=&-\frac{i}{H^4}\int_{z_\Lambda}^{\infty} dz \Delta m^2(T)\frac{k^3}{z^4}v_k^2\\
	|\xi|&=&\frac{1}{H^4}\int_{z_\Lambda}^{\infty} dz \Delta m^2(T)\frac{k^3}{z^4}|v_k|^2~.
	\end{eqnarray}
	Now use Baker-Campbell-Hausdorff formula again. Split the exponential into two parts:
	\begin{eqnarray}
	\nonumber S&=&\exp\left[\frac{1}{2}\int\frac{d^3k}{(2\pi)^3}\left( \varpi a_\mathbf{k}a_\mathbf{-k}-\varpi^{*}a_{\mathbf{k}}^\dagger a_{-\mathbf{k}}^\dagger\right)\right]\exp\left[-\frac{i}{2}\int\frac{d^3k}{(2\pi)^3}|\xi|\left(a_\mathbf{k}a_\mathbf{k}^\dagger+a_\mathbf{k}^\dagger a_\mathbf{k}\right)\right]\times(1+\mathcal{O}(\Delta m^4))\\
	&\equiv&S_1S_2\times(1+\mathcal{O}(\Delta m^4))~.
	\end{eqnarray}
	After imposing normal ordering, the $S_2$ operator is nothing other than the particle-number operator exponentiated. The effect of $S_2$ on the BD vacuum is just a multiplication by the identity. Therefore it is the first factor $S_1$ that changed the structure of the BD vacuum. Writing out $S_1$ explicitly, we see that it is exactly the squeezing operator of a squeezing parameter $\varpi$.
	\begin{equation}
	|z_\Lambda\rangle=S|\Omega\rangle\approx S_1|\Omega\rangle=\exp\left[\frac{1}{2}\int\frac{d^3k}{(2\pi)^3}\left( \varpi a_\mathbf{k}a_\mathbf{-k}-\varpi^{*}a_{\mathbf{k}}^\dagger a_{-\mathbf{k}}^\dagger\right)\right]|\Omega\rangle~.
	\end{equation}
	The consequence of squeezing the BD vacuum is a squeezed state with a spectrum of particles in it. And it is related to the BD vacuum by a Bogolyubov transformation. In other words, the state $|z_\Lambda\rangle$ corresponds to a non-BD state,
	\begin{eqnarray}
	\nonumber S a_\mathbf{p}S^\dagger&=&\tilde{a}_\mathbf{p}=a_\mathbf{p}\cosh r +a_\mathbf{-p}^\dagger e^{i\theta}\sinh r \\
	\text{for}~~~S&=&\exp\left[\frac{1}{2}\int\frac{d^3k}{(2\pi)^3}r\left( e^{-i\theta} a_\mathbf{k}a_\mathbf{-k}-e^{i\theta}a_{\mathbf{k}}^\dagger a_{-\mathbf{k}}^\dagger\right)\right]~.
	\end{eqnarray}
	In our case, 
	\begin{equation}\label{nonBDcoeff}
	\varpi=r e^{-i\theta}\approx-\frac{\Delta m^2(T)}{4H^2 z_\Lambda^2}e^{2iz_\Lambda}~.
	\end{equation}
	for large $z_\Lambda$. So the number density of massive particles produced by the UV furnace is
	\begin{equation}\label{rapprox}
	\langle z_\Lambda|a_\mathbf{p}^\dagger a_\mathbf{p}|z_\Lambda\rangle=\sinh^2 r\approx \left(\frac{\Delta m^2(T)}{4H^2z_\Lambda^2}\right)^2~.
	\end{equation}
	Notice that the spectrum is independent of the comoving momentum. This is reasonable because the comoving momentum is unphysical and can be rescaled to any value by a redefinition of the space coordinates. 
	
	One important comment. In \cite{Jiang:2016nok}, it has been proved that field excitations with a non-BD initial condition at $\tau=-\infty$ decay quickly through the inflaton loop, which shows exactly why BD initial condition is a more physical choice in the conventional inflation. The decay process is described by
	\begin{equation}
	\text{(Non-BD coefficient)}\propto e^{-\Gamma (\tau-\tau_\Lambda)},~~\Gamma\sim f_{NL}^2 P_\zeta k^3\tau_\Lambda^2
	\end{equation}
	for a dimension-five operator $\mathcal{O}_5$. In our case, however, we do have a non-BD initial condition but the starting point is not infinity but a certain scale $z_\Lambda=-k\tau_\Lambda$. The characteristic time scale is $\Gamma^{-1}\sim f_{NL}^{-2} P_\zeta^{-1} k^{-3}\tau_\Lambda^{-2}$. For a moderate choice of parameters, $P_\zeta\sim 10^{-9},~f_{NL}\sim 1, z_\Lambda\sim 10^3$, the decay factor is $\sim e^{-1}\sim 0.4$. So the non-BD state actually did not decay much. We can absorb the weakening effect into the definition of squeezing parameter $r$. Therefore the non-BD modes can survive long enough to leave footprints on the correlation functions. 
	
\begin{figure}[htbp] 
		\centering 
		\includegraphics[width=15cm]{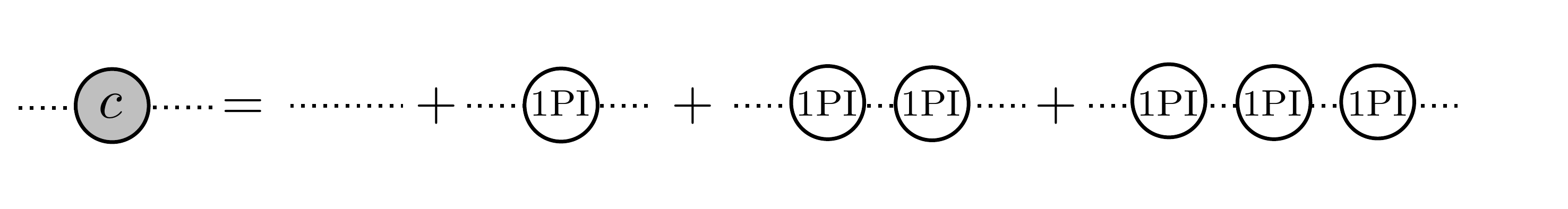}   
		\caption{An alternative understanding of the origin of the effective vertex. C for connected.} 
		\label{FeynSum}
	\end{figure}

	Finally we note that our formalism is equivalent of a summation of all the one loop diagrams in the UV thermal equilibrium. And the approximation $\sinh r\approx r$ in (\ref{rapprox}) is essentially the leading order in FIG.~\ref{FeynSum}.
	
	Apart from using a non-BD initial condition at a cutoff scale, it is also possible to regard (\ref{interactionhamiltonian}) as an operator in the UV and use the conventional in-in formalism to calculate the observables on the boundary $\tau=0$. We give a example of such calculations in Appendix.~\ref{AppendixC}. The results are the same as those obtained under the non-BD initial state. With the non-BD initial state, it is natural to expect terms with no Boltzmann suppression factor of the form $\exp(-\pi \mu)$ in the standard clock signal for particles with masses $H<m<H z_\Lambda<T$.
	
	\section{A Quantum Mechanics problem}\label{QMsection}
	To solidify the formal constructions above, we give another way to understand the effect brought about by the abrupt exit from the thermal equilibrium and thus the sudden change of mass. It can be formulated in a way that strongly resembles the scattering of a wave by a unit step potential in quantum mechanics. 
	\begin{figure}[htbp] 
		\centering 
		\includegraphics[width=9cm]{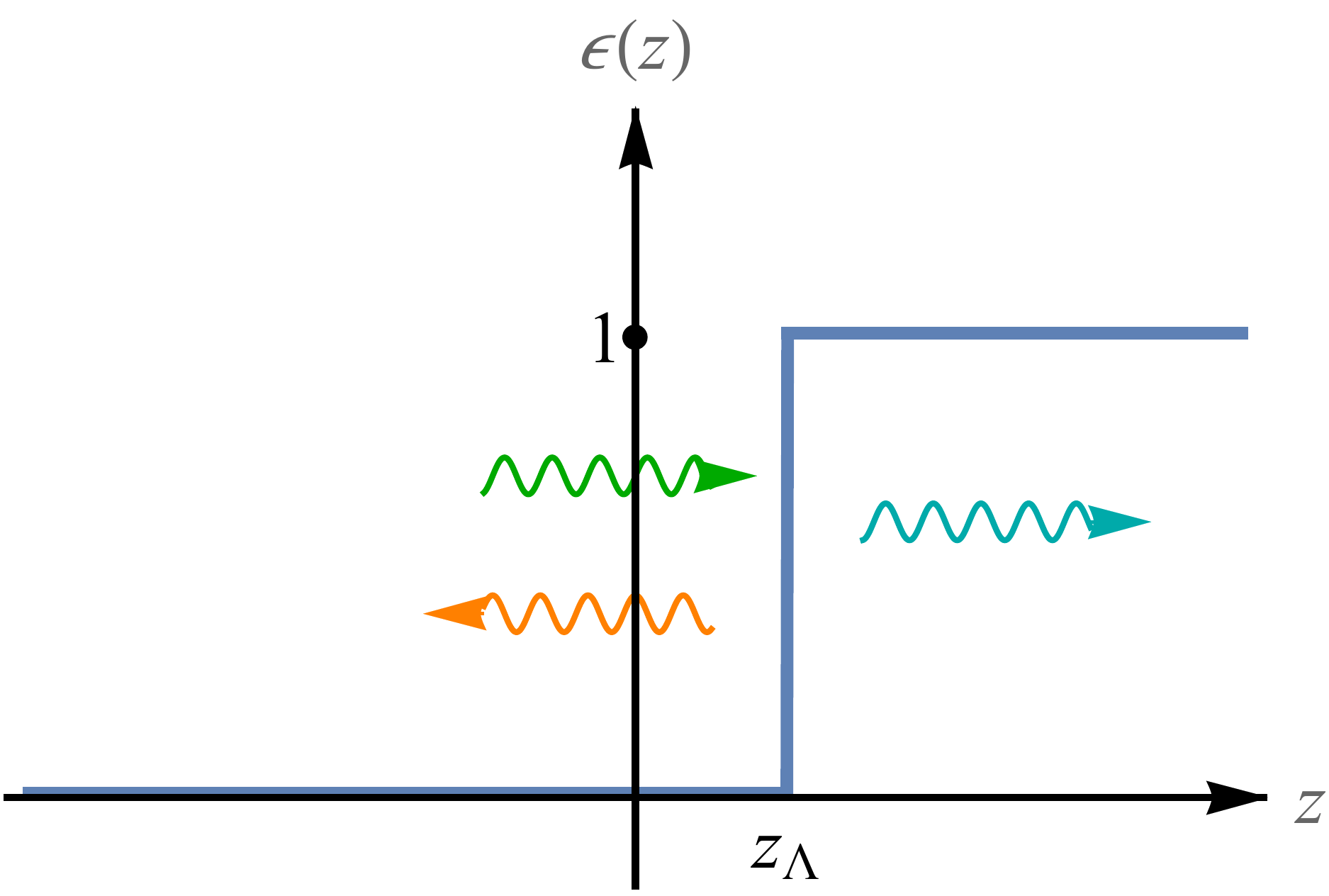}   
		\caption{A Quantum Mechanics problem: scattering off a unit step potential} 
		\label{QMScattering}
	\end{figure}
	
	The wave functions are constructed from basic solutions in two regions of the z axis and are sewn together at the point $z_\Lambda$ so that the wave function and its first order derivative stay continuous across $z_\Lambda$. The wave function in the right region is a  positive-frequency mode function with mass squared $m^2+\Delta m^2$ while the one in the left region is a combination of positive- and negative-frequency mode functions with mass squared $m^2$. The coefficients are nothing other than the Bogolyubov coefficients,
	\begin{eqnarray}
	v_k(\tau_\Lambda^-)|_{m^2+\Delta m^2}&=&\alpha v_k(\tau_\Lambda^+)|_{m^2}+\beta v_k(\tau_\Lambda^+)^*|_{m^2}\\
	v_k'(\tau_\Lambda^-)|_{m^2+\Delta m^2}&=&\alpha v_k'(\tau_\Lambda^+)|_{m^2}+\beta v_k'(\tau_\Lambda^+)^*|_{m^2}~.
	\end{eqnarray}
	Plug in the analytic expression (\ref{modefun}) and solve for the coefficients up to lowest order gives
	\begin{equation}
	\alpha\approx 1-\frac{i\Delta m^2(T)}{2Hz_\Lambda}+\mathcal{O}\left(\frac{\Delta m^4(T)}{H^2z_\Lambda^2}\right),~\beta=\frac{\Delta m^2(T)}{4H^2z_\Lambda^2}e^{2iz_\Lambda}+\mathcal{O}\left(\frac{\Delta m^2(T)}{H^2z_\Lambda^3}\right)~.
	\end{equation}
	Comparing to (\ref{nonBDcoeff}), we see that these two approaches give identical results. Physically, however, the process of exiting the thermal equilibrium is not an abrupt process. Instead, it should happen gradually. Hence the shape of the potential barrier is not a unit step but a more smooth one. A straightforward matching method above is applicable against a simple unit step potential but is incapable of solving complicated potentials, for which we need to solve perturbatively.
	
\begin{figure}[htbp] 
		\centering 
		\includegraphics[width=14cm]{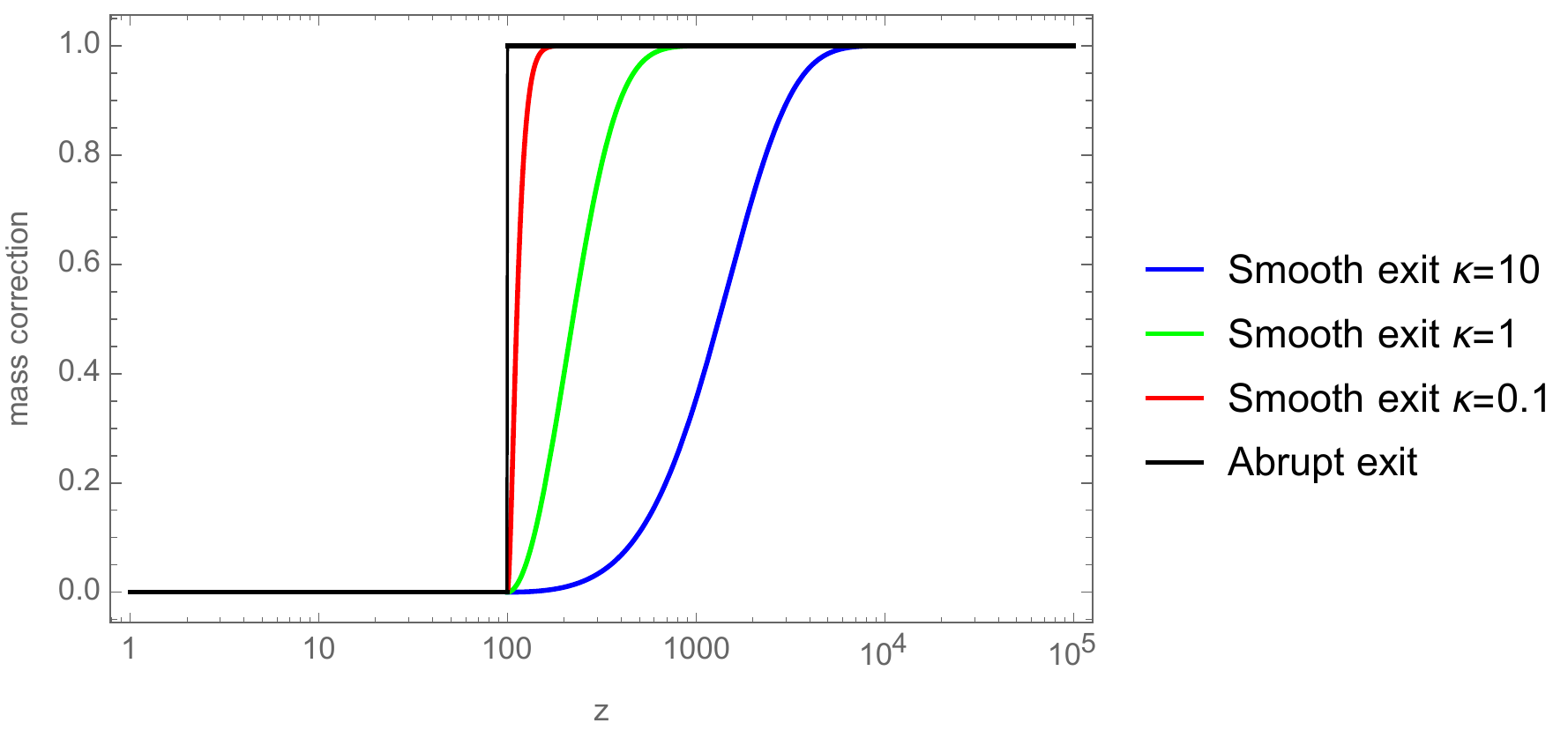}   
		\caption{An illustration of the function $\epsilon(z)$. The black line denotes the effective two-point vertex has an abrupt exit. The blue, green and red line is the smooth exit case where the parameter $\kappa=10,1,0.1$ respectively. We can see that as $\kappa$ becomes smaller, the smooth exit case tends to the abrupt exit case.} \label{ExitStyle}
	\end{figure}

	Let us consider a mass correction term evolving in time with some function $\epsilon(\tau)$.
	\begin{equation}\label{EFVoperator}
	\mathcal{\tilde{H}}_I\supset \frac{1}{2}\Delta m^2(T) a^4\sigma^2\epsilon(\tau)~.
	\end{equation}
	We would like to estimate the squeezed parameter $\varpi$ in this case. We choose the following parametrization of the $\epsilon$ function to describe the time dependence of the two-point vertex,
	\begin{align}
	\epsilon(z)=(1-e^{-(z/z_\Lambda-1)/\kappa})^2\theta(z-z_\Lambda)~.
	\end{align} 
	The parameter $\kappa$ parametrize the rate of the coupling decreasing to zero. As $\kappa\rightarrow 0$, this function tends to the Heaviside theta function. An illustration of this function is shown in FIG.~\ref{ExitStyle}.
	
	Starting from (\ref{varpi}), with $\Delta m^2(T)$ replaced by $\Delta m^2(T,z)=\Delta m^2(T)\epsilon(z)$, the integral can be evaluated as
	\begin{align}\nonumber
	&\int_{z_\Lambda}^{\infty} z^{-2} e^{2iz} (1-e^{-(z/z_\Lambda-1)/\kappa})^2dz \\\nonumber
	&= \frac{2 e^{\frac{1}{\kappa }} }{\kappa  z_{\Lambda }}\left(\left(1-2 i \kappa  z_{\Lambda }\right) \Gamma \left(0,\frac{1}{\kappa }-2 i z_{\Lambda }\right)+i e^{\frac{1}{\kappa }}
	\left(\kappa  z_{\Lambda }+i\right) \Gamma \left(0,\frac{2}{\kappa }-2 i z_{\Lambda }\right)\right)+2 i \Gamma \left(0,-2 i z_{\Lambda
	}\right) \\
	& = \left\{
	\begin{aligned}
	&-\frac{i e^{2iz_\Lambda}}{4 z_\Lambda^4\kappa^2} + \mathcal{O} (z_\Lambda^{-5}),~\text{expansion at large $z_\Lambda$.}\\
	&\frac{i e^{2iz_\Lambda}}{2 z_\Lambda^2} + \mathcal{O} (\kappa^2,z_\Lambda^{-1}),~\text{expansion first at $\kappa=0$ and then at large $z_\Lambda$.}
	\end{aligned}\right.
	\end{align}
	The first scenario corresponds to a smooth exit from the thermal equilibrium and the second one returns to the abrupt exit limit. Making use of \eqref{varpi}, we can obtain 
	\begin{eqnarray}
	r e^{-i\theta}=\varpi=\left\{
	\begin{aligned} & \frac{\Delta m^2(T)}{8 H^2 z_\Lambda^4\kappa^2}e^{2iz_\Lambda},~\text{a smooth exit with exit parameter $\kappa$.}\\
	&-\frac{\Delta m^2(T)}{4 H^2 z_\Lambda^2}e^{2iz_\Lambda},~\text{an abrupt exit.}
	\end{aligned}\right.~.\label{UVvarpi}
	\end{eqnarray}
	We move on to consider the WKB condition of the abrupt exit and the smooth exit. We get $\dot w/w^2<1$ for the smooth exit case. It means the WKB condition is not violated and the particle production is still subject to the exponential suppression but changing from $\exp(-\pi\mu)$ to $1/(\exp(-z_\Lambda H/T)-1)$. There are two changes here. One is the energy changes from $m$ to $z_\Lambda$ essentially because in the conventional cold quasi-single field inflation, particle production is at IR and the energy is dominated by mass. Here the particle production is mainly in the UV region and energy is dominated by $-k\tau$ which is roughly $z_\Lambda$. Although the energy needed to produce a particle is larger, the temperature is also higher compared with the Hubble temperature. So if $m/H>z_\Lambda H/T$, we can have an enhanced clock signal compared to the original cold quasi-single field inflation. For the abrupt exit case, we got the WKB violation $\dot w/w^2>1$ at the point $z=z_\Lambda$ because of the Heaviside function. But it does not imply a large amount of particles are produced because the change of mass is assumed to be small. From our analysis above, the particle production rate in this case is still subject to suppression as in the smooth exit case.
	
	\section{The three-point function}\label{3ptsection}
	In this section, we will put our non-BD initial condition to use and calculate the three-point function explicitly. There are two equivalent formalisms developed for the calculation of observables during inflation, namely the in-in formalism and the Schwinger-Keldysh path integral formalism. We used the in-in expression in (\ref{ininobservable}) since we were dealing with quantum states and in-in formalism is more suitable. However, it is easier to do actual computations of correlating functions in Schwinger-Keldysh formalism. Henceforth we will switch to this formalism. As a double check, another equivalent calculation using in-in formalism and the effective vertex method is presented in Appendix.~\ref{AppendixC}.
	
	Before calculation, let us first analyze the possible terms and their contribution to the clock signal. To track the positive frequency part, we assign an arrow to each propagator to indicate the propagating direction of their positive frequency part. The arrow always points from $v^*$ to $v$.
	
	\begin{figure}[htbp] 
		\centering 
		\includegraphics[width=10cm]{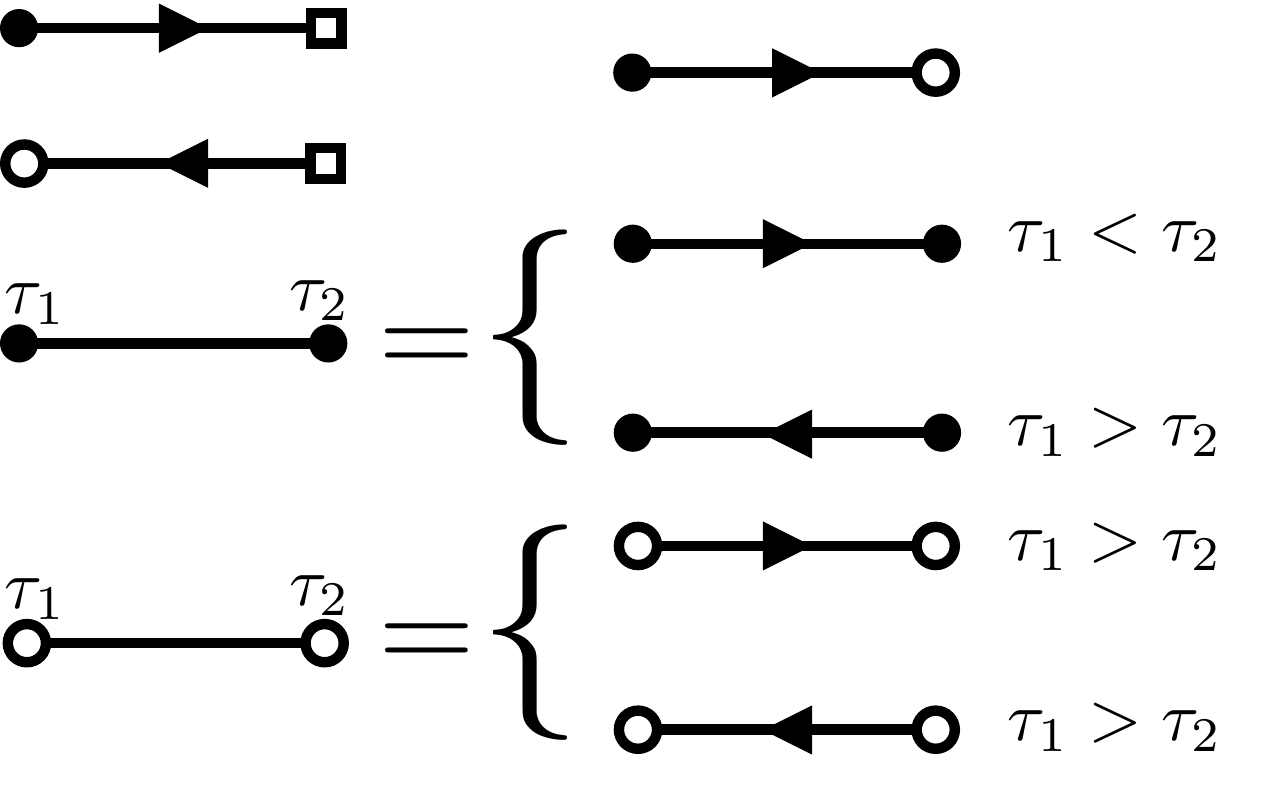}   
		\caption{Diagrammatic arrow notation. Black bots are for $+$ and white dots for $-$.} 
	\end{figure}
	
\begin{figure}[htbp] 
		\centering 
		\includegraphics[width=10cm]{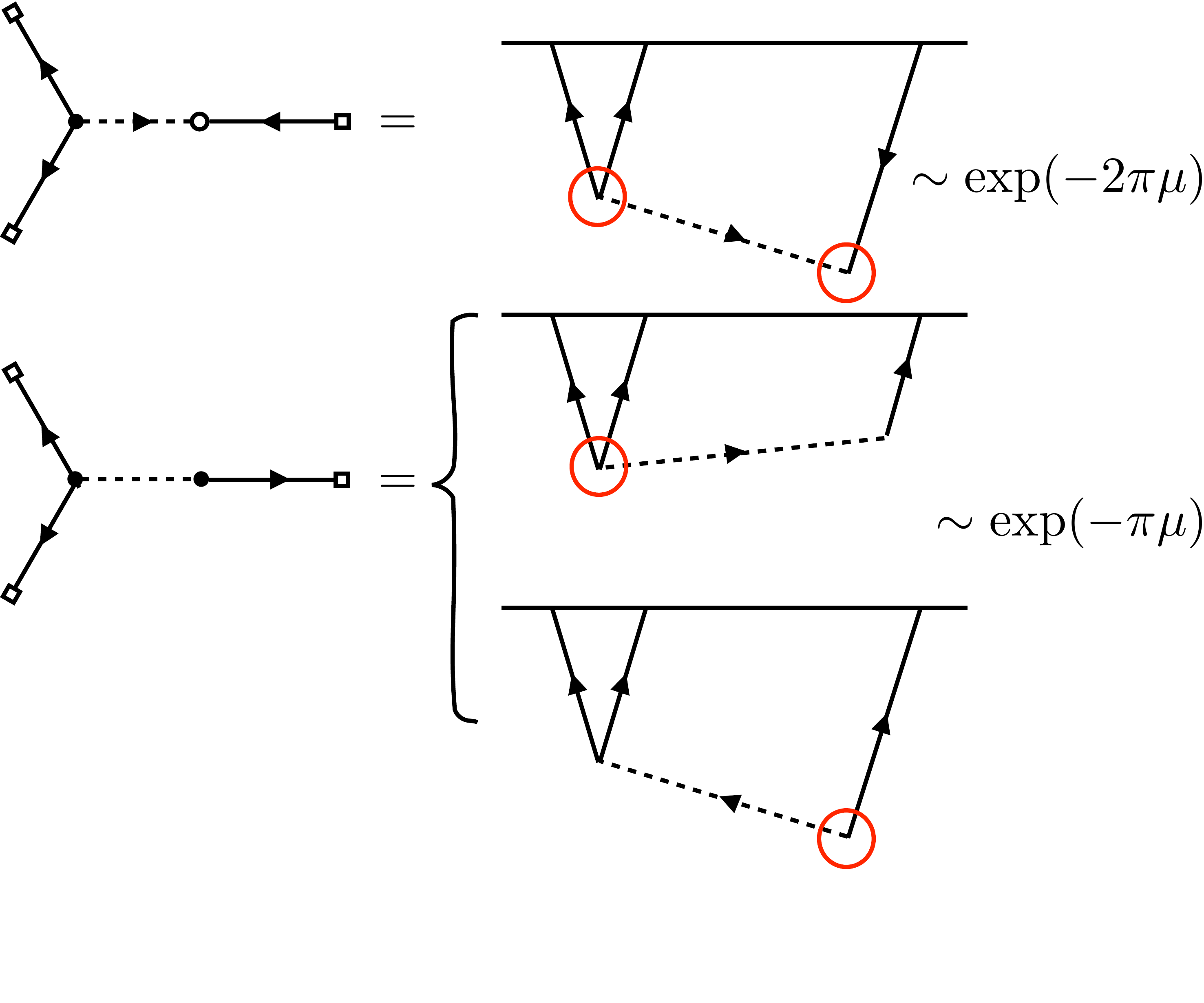}   
		\caption{Diagrammatic analysis of the conventional cold quasi-single field inflation three-point diagram.} \label{CQSFIArrows}
	\end{figure}

	In this notation we can easily see why the conventional cold quasi-single field inflation only gives Boltzmann-suppressed clock signals. In FIG.\ref{CQSFIArrows}, the diagram without time ordering has two vertices where arrows pop out of nowhere or disappear into nothing. This indicates that to achieve resonance, we need one negative frequency part from each of the two vertices. Without resonance and an explicit cutoff, the integral oscillates very fast with respect to the integration upper limit and gives effectively zero contribution. Therefore the non-timed-ordered diagram is doubly suppressed by $\exp(-2\pi \mu)$. The time-ordered diagram, however, only receives one $\exp(-\pi \mu)$ suppression factor. The other two diagrams are similar since they are given by complex conjugation.
	
\begin{figure}[h!] 
		\centering 
		\includegraphics[width=15cm]{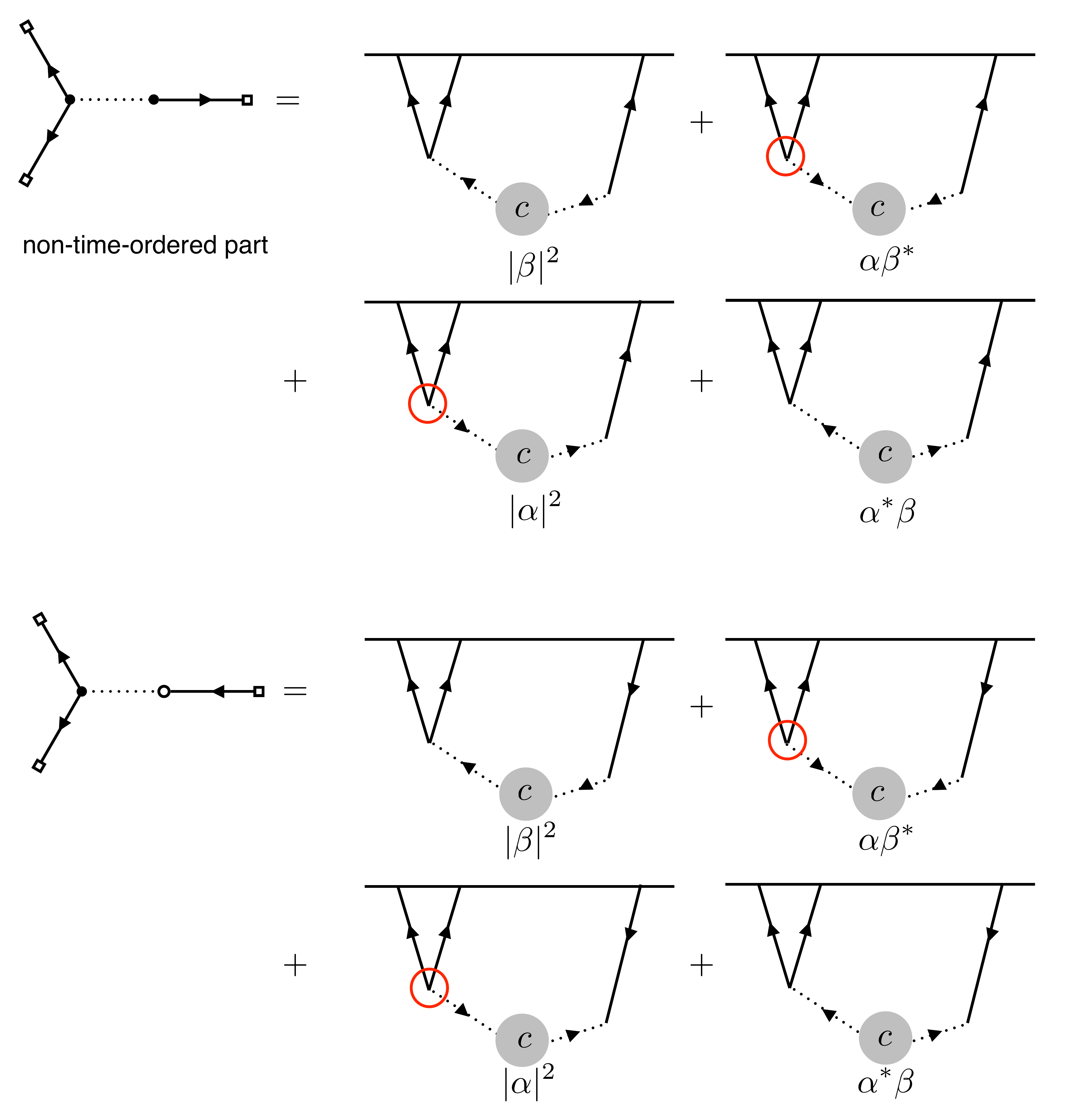}   
		\caption{Diagrammatic analysis of warm quasi-single field inflation three-point diagram. Notice now the propagator of massive fields are replaced by non-BD ones and so we changed dashed lines to dotted lines. In each diagram there are two subdiagram containing unsuppressed clock signals in the sense that the triple vertex be on-shell. These four subdiagrams correspond to the first two orders in the squeezed limit of the bispectrum, $ B_\phi(k,k,ck)^{(1)}$ in (\ref{clock1}) and $ B_\phi(k,k,ck)^{(2)}$ in (\ref{clock2}).} \label{WQSFIArrows}
	\end{figure}

	Now in warm quasi-single field inflation, as soon as non-BD mode functions are applied, things become much different. The massive field propagator is built from a new non-BD mode function $\tilde{v}_k(\tau)$ related to the initial ones by a Bogolyubov transformation,
	\begin{equation}
	\tilde{v}_k=\alpha v_k+\beta v_k^*~,~~~|\alpha|^2-|\beta|^2=1.
	\end{equation}
	The corresponding annihilation operator is
	\begin{equation}\label{BogolyubovCoeff}
	\tilde{a}_\mathbf{p}=\alpha^* a_\mathbf{p} - \beta^* a_\mathbf{-p}^\dagger~,~~~\tilde{a}_\mathbf{p}|z_\Lambda\rangle=\tilde{a}_\mathbf{p}S|\Omega\rangle=S^\dagger a_\mathbf{p}|\Omega\rangle=0~,~~~\alpha=\cosh r~,~\beta=-e^{-i\theta}\sinh r~.
	\end{equation} 
	where $v_k$ is defined in (\ref{modefun}). Therefore the non-BD propagator for the massive field contains four parts, of which three are strongly dependent on the integration upper limit (i.e. the cutoff) indicated in FIG.~\ref{WQSFIArrows}. Naively one would guess that these three parts give zero contribution to the final unsuppressed clock signal. However, a closer look reveals that our UV results (\ref{UVvarpi}) are also dependent on the cutoff $z_\Lambda$. It is possible that these two dependences may cancel each other out and gives a result less dependent on the cutoff. We will demonstrate this cancellation in the following calculation. 
	
	Now we move on to analytical computation. In the Schwinger-Keldysh formalism, there are four different massive field Green functions
	\begin{eqnarray}
	D_{++}(k,\tau_1,\tau_2)&=&\theta(\tau_1-\tau_2)\tilde v_k(\tau_1)\tilde v_k(\tau_2)^*+\theta(\tau_2-\tau_1)\tilde v_k(\tau_1)^*\tilde v_k(\tau_2)\\
	D_{+-}(k,\tau_1,\tau_2)&=&\tilde v_k(\tau_1)^*\tilde v_k(\tau_2)\\
	D_{-+}(k,\tau_1,\tau_2)&=&\tilde v_k(\tau_1)\tilde v_k(\tau_2)^*\\
	D_{--}(k,\tau_1,\tau_2)&=&\theta(\tau_1-\tau_2)\tilde v_k(\tau_1)^*\tilde v_k(\tau_2)+\theta(\tau_2-\tau_1)\tilde v_k(\tau_1)\tilde v_k(\tau_2)^*~,
	\end{eqnarray}
	where the mode function is given by (\ref{modefun}). The inflaton propagators are denoted by $G$ and are obtained by setting $m=0$. Using the Feynman rules derived in \cite{Chen:2017ryl}, the diagram reads
	\begin{eqnarray}
	\nonumber\langle \phi_\mathbf{k1}\phi_\mathbf{k2}\phi_\mathbf{k3}\rangle'&\equiv& B_\phi(k_1,k_2,k_3)=\int \frac{d\tau_1}{(-H\tau_1)^2} \frac{d\tau_2}{(-H\tau_2)^3}C_2 C_3\\
	&&\times\left((-1)^{a+b}\partial_{\tau_1}G_{a,1-a}(k_1,\tau_1,0)\partial_{\tau_1}G_{a,1-a}(k_2,\tau_1,0)D_{ab}(k_3,\tau_1,\tau_2)\partial_{\tau_2}G_{b,1-b}(k_3,\tau_2,0)\right),~~~~
	\end{eqnarray}
	where summation over $a,b=\{1(+),0(-)\}$ is assumed. The primed expectation value is related to the $n$-point function through $\langle Q\rangle=(2\pi)^3\delta^3(\sum_i \mathbf{k}_i)\langle Q\rangle'$. 
	
	If we proceed in the standard way and insert the mode functions, we will arrive at a very long expression containing all the contributions drawn in FIG.~\ref{WQSFIArrows}. Then we set $\tau_1>\tau_2$ and use the theta function property to convert part of the time-ordered terms into non-time-ordered terms. The residual time-ordered  terms are of order $|\alpha|^2$ and $|\beta|^2$.
	\begin{eqnarray}
	\nonumber && |\alpha| ^2 \left[\frac{\partial u_{k_1}\left(\tau _1\right){}^*}{\partial \tau _1} \frac{\partial u_{k_2}\left(\tau _1\right){}^*}{\partial
		\tau _1} \frac{\partial u_{k_3}\left(\tau _2\right){}^*}{\partial \tau _2} u_{k_1}(0) u_{k_2}(0) u_{k_3}(0) v_{k_3}\left(\tau _2\right) v_{k_3}\left(\tau _1\right){}^*\right.\\
	\nonumber&&\left.-\frac{\partial
		u_{k_3}\left(\tau _2\right)}{\partial \tau _2} \frac{\partial u_{k_1}\left(\tau _1\right){}^*}{\partial \tau _1} \frac{\partial u_{k_2}\left(\tau _1\right){}^*}{\partial \tau _1}
	u_{k_1}(0) u_{k_2}(0) u_{k_3}(0){}^* v_{k_3}\left(\tau _2\right) v_{k_3}\left(\tau _1\right){}^*\right]\\
	\nonumber	&&+\alpha ^* \beta 
	\left[\frac{\partial u_{k_1}\left(\tau _1\right){}^*}{\partial \tau _1} \frac{\partial u_{k_2}\left(\tau _1\right){}^*}{\partial \tau _1} \frac{\partial u_{k_3}\left(\tau
		_2\right){}^*}{\partial \tau _2} u_{k_1}(0) u_{k_2}(0) u_{k_3}(0) v_{k_3}\left(\tau _1\right){}^* v_{k_3}\left(\tau _2\right){}^*\right.\\
	\nonumber&&\left.-\frac{\partial u_{k_1}\left(\tau _1\right)}{\partial
		\tau _1} \frac{\partial u_{k_2}\left(\tau _1\right)}{\partial \tau _1} \frac{\partial u_{k_3}\left(\tau _2\right){}^*}{\partial \tau _2} u_{k_3}(0) u_{k_1}(0){}^* u_{k_2}(0){}^*
	v_{k_3}\left(\tau _1\right){}^* v_{k_3}\left(\tau _2\right){}^*\right.\\
	\nonumber&&\left.-\frac{\partial u_{k_3}\left(\tau _2\right)}{\partial \tau _2} \frac{\partial u_{k_1}\left(\tau _1\right){}^*}{\partial
		\tau _1} \frac{\partial u_{k_2}\left(\tau _1\right){}^*}{\partial \tau _1} u_{k_1}(0) u_{k_2}(0) u_{k_3}(0){}^* v_{k_3}\left(\tau _1\right){}^* v_{k_3}\left(\tau
	_2\right){}^*\right.\\
	\nonumber&&\left.+\frac{\partial u_{k_1}\left(\tau _1\right)}{\partial \tau _1} \frac{\partial u_{k_2}\left(\tau _1\right)}{\partial \tau _1} \frac{\partial u_{k_3}\left(\tau
		_2\right)}{\partial \tau _2} u_{k_1}(0){}^* u_{k_2}(0){}^* u_{k_3}(0){}^* v_{k_3}\left(\tau _1\right){}^* v_{k_3}\left(\tau _2\right){}^*\right]\\
	\nonumber	&&+|\beta|^2 \left[\frac{\partial
		u_{k_1}\left(\tau _1\right){}^*}{\partial \tau _1} \frac{\partial u_{k_2}\left(\tau _1\right){}^*}{\partial \tau _1} \frac{\partial u_{k_3}\left(\tau _2\right){}^*}{\partial \tau _2}
	u_{k_1}(0) u_{k_2}(0) u_{k_3}(0) v_{k_3}\left(\tau _1\right) v_{k_3}\left(\tau _2\right){}^*\right.\\
	\nonumber&&\left.-\frac{\partial u_{k_3}\left(\tau _2\right)}{\partial \tau _2} \frac{\partial
		u_{k_1}\left(\tau _1\right){}^*}{\partial \tau _1} \frac{\partial u_{k_2}\left(\tau _1\right){}^*}{\partial \tau _1} u_{k_1}(0) u_{k_2}(0) u_{k_3}(0){}^* v_{k_3}\left(\tau _1\right)
	v_{k_3}\left(\tau _2\right){}^*\right]\\
	\nonumber&&+\Theta \left(\tau _1-\tau _2\right)\\
	\nonumber&&\times\left[ -\frac{\partial u_{k_1}\left(\tau _1\right){}^*}{\partial \tau _1} \frac{\partial u_{k_2}\left(\tau _1\right){}^*}{\partial \tau _1} \frac{\partial
		u_{k_3}\left(\tau _2\right){}^*}{\partial \tau _2} u_{k_1}(0) u_{k_2}(0) u_{k_3}(0) v_{k_3}\left(\tau _2\right) v_{k_3}\left(\tau _1\right){}^*\right.\\
	&&\left.+\frac{\partial u_{k_1}\left(\tau _1\right){}^*}{\partial \tau _1} \frac{\partial u_{k_2}\left(\tau
		_1\right){}^*}{\partial \tau _1} \frac{\partial u_{k_3}\left(\tau _2\right){}^*}{\partial \tau _2} u_{k_1}(0) u_{k_2}(0) u_{k_3}(0) v_{k_3}\left(\tau _1\right) v_{k_3}\left(\tau
	_2\right){}^*\right]+\text{c.c.}\label{rawclock}~,
	\end{eqnarray}
	where in the time-ordered part the identity $|\alpha|^2-|\beta|^2=1$ is applied.
	
	\subsection{non-time-ordered part}
	Before proceeding, we recall that the oscillatory behavior in the squeezed limit is due to a resonance effect between the inflaton and the massive field. The inflaton oscillates in the conformal time $\tau$ before exiting Hubble horizon while the massive fields oscillate in the physical time $t$, as can be seen in the IR approximation of its mode function,
	\begin{equation}\label{IRexpansion}
	v_k(\tau)\xrightarrow{-k\tau\ll \mu}\frac{(1-i)   2^{-\frac{3}{2}-i \mu }\sqrt{\pi } e^{-\frac{\pi  \mu }{2}}
		(\coth (\pi  \mu )+1)H }{k^{3/2} \Gamma (i
		\mu +1)}(-k \tau )^{\frac{3}{2}+i \mu }-\frac{(1+i) 2^{-\frac{3}{2}+i \mu } e^{-\frac{\pi  \mu }{2}}
		\Gamma (i \mu )H }{\sqrt{\pi } k^{3/2}}(-k \tau )^{\frac{3}{2}-i \mu }~.
	\end{equation}
	The clock signal appears in the squeezed limit of the bispectrum. Therefore we set $k_1=k_2=k,~k_3=ck,~c\ll 1$. For the non-time-ordered part, the resonance happens when the integrand in
	\begin{equation}
	\int_{\tau_\Lambda} d\tau u_k'^{*2}v_{ck}\sim \int_{\tau_\Lambda} d\tau e^{2ik\tau}H^{(1)}_{i\mu}(-ck\tau)\sim c^{i\mu}\int_{\tau_\Lambda} d\tau e^{2ik\tau+i\mu \ln(-H\tau)}\approx c^{i\mu}\int_{\tau_\Lambda} d\tau e^{2ik\tau-imt}~.
	\end{equation}
	encounters a saddle point at $2k\tau_*+\mu \ln(-H\tau_*)=2k\tau_*-mt_*=0$. The resonance renders the integral to be independent on the cutoff $\tau_\Lambda=-z_\Lambda/c k$. On the other hand, if the vertex integral is off-shell, the integral will strongly depend on the cutoff,
	\begin{equation}
	\int_{\tau_\Lambda} d\tau u_k'^{*2}v^*_{ck}\sim \int_{\tau_\Lambda} d\tau e^{2ik\tau}H^{(2)}_{-i\mu}(-ck\tau)\sim c^{-i\mu}\int_{\tau_\Lambda} d\tau e^{2ik\tau-i\mu \ln(-H\tau)}\approx c^{-i\mu}\int_{\tau_\Lambda} d\tau e^{2ik\tau+imt}\propto c^{-i\mu} e^{-\frac{2iz_\Lambda}{c}}~.
	\end{equation}
	The cutoff $z_\Lambda \gg1$ of the IR integral is usually rather large and the squeezed limit $c\ll 1$ is small. Hence the off-shell vertex integral oscillates extremely fast with respect to $c$. And the corresponding signatures on the bispectrum are coarse-grained out and are practically invisible. As a result, we only focus on the diagrams with the $\phi'^2\sigma$ vertex on shell. A diagrammatic illustration is shown in FIG.~\ref{WQSFIArrows}, where off-shell triple vertices are labeled. We throw away the corresponding terms in the non-time-ordered integrand. As for the time-ordered part, later in the next subsection, we will prove that there are no unsuppressed clock signals in the non-time-ordered part. The terms left are
	\begin{eqnarray}
	\nonumber
	\text{(integrand)}&=&\alpha ^* \beta  \left(-\frac{\partial u_{k_1}\left(\tau _1\right)}{\partial \tau _1} \frac{\partial u_{k_2}\left(\tau _1\right)}{\partial \tau _1} \frac{\partial
		u_{k_3}\left(\tau _2\right){}^*}{\partial \tau _2} u_{k_3}(0) u_{k_1}(0){}^* u_{k_2}(0){}^* v_{k_3}\left(\tau _1\right){}^* v_{k_3}\left(\tau _2\right){}^*\right.\\ \label{3ptalphabeta}
	&&\left.+\frac{\partial
		u_{k_1}\left(\tau _1\right)}{\partial \tau _1} \frac{\partial u_{k_2}\left(\tau _1\right)}{\partial \tau _1} \frac{\partial u_{k_3}\left(\tau _2\right)}{\partial \tau _2}
	u_{k_1}(0){}^* u_{k_2}(0){}^* u_{k_3}(0){}^* v_{k_3}\left(\tau _1\right){}^* v_{k_3}\left(\tau _2\right){}^*\right)\\ \label{3ptbetabeta}
	\nonumber&&+|\beta| ^2 \left(\frac{\partial u_{k_1}\left(\tau
		_1\right){}^*}{\partial \tau _1} \frac{\partial u_{k_2}\left(\tau _1\right){}^*}{\partial \tau _1} \frac{\partial u_{k_3}\left(\tau _2\right){}^*}{\partial \tau _2} u_{k_1}(0)
	u_{k_2}(0) u_{k_3}(0) v_{k_3}\left(\tau _1\right) v_{k_3}\left(\tau _2\right){}^*\right.\\
	&&\left.-\frac{\partial u_{k_3}\left(\tau _2\right)}{\partial \tau _2} \frac{\partial u_{k_1}\left(\tau
		_1\right){}^*}{\partial \tau _1} \frac{\partial u_{k_2}\left(\tau _1\right){}^*}{\partial \tau _1} u_{k_1}(0) u_{k_2}(0) u_{k_3}(0){}^* v_{k_3}\left(\tau _1\right) v_{k_3}\left(\tau
	_2\right){}^*\right)\\
	&&+\text{(time-ordered part)}+\text{c.c.}~.
	\end{eqnarray}
	
	Perform two integrations, we obtain
	\begin{equation}
	B_\phi(k,k,ck)=C_2C_3\int_{-z_\Lambda/(ck)}^{0}\frac{d\tau_1}{(-H\tau_1)^2}\left[\int_{-z_\Lambda/(ck)}^{0}\frac{d\tau_2}{(-H\tau_2)^3}\times\text{(integrand)}\right]_{c\rightarrow 0}~.
	\end{equation}
	The final result for the bispectrum is
	\begin{eqnarray}
	B_\phi(k,k,ck)&=&\alpha^* \beta B_\phi(k,k,ck)^{(1)}+|\beta|^2 B_\phi(k,k,ck)^{(2)}+\text{c.c.}~,
	\end{eqnarray}
	where the first and second order of unsuppressed clock signals are
	\begin{eqnarray}
	\label{clock1} B_\phi(k,k,ck)^{(1)}&=&C_2 C_3 \frac{H^3\mu ^{3/2}}{1024k^6}\sqrt{\frac{\pi}{2} } \left(\frac{c}{4}\right)^{-\frac{3}{2}-i \mu }\left[\frac{1-i}{z_{\Lambda }}+e^{2iz_\Lambda}(1+i) \left(\log \left(\frac{4 z_{\Lambda }^2}{\mu ^4}\right)+i \pi-2 \gamma \right)\right]\\
	\label{clock2} B_\phi(k,k,ck)^{(2)}&=&C_2 C_3 \frac{H^3\mu ^{3/2}}{1024k^6}\sqrt{\frac{\pi}{2}}\left(\frac{c}{4}\right)^{-\frac{3}{2}-i \mu }\left[(1+i)  \left(\log \left(\frac{4 z_{\Lambda }^2}{\mu ^4}\right)-i \pi -2 \gamma
	\right)-\frac{(1-i) e^{2 i z_{\Lambda }}}{z_{\Lambda }}\right]~.~~
	\end{eqnarray}
	During the calculation, some approximations based on the hierarchy $c z_\Lambda< \mu\ll z_\Lambda\ll z_\Lambda/c$ are used. Remember our Bogolyubov coefficients are fixed by the UV physic to be $\alpha=\cosh r,~\beta=-e^{-i\theta}\sinh r$. Therefore,
	\begin{eqnarray}
	B_\phi(k,k,ck)&=&e^{2iz_\Lambda}\cosh r\sinh r B_\phi(k,k,ck)^{(1)}+\sinh^2 r B_\phi(k,k,ck)^{(2)}+\text{c.c.}~,
	\end{eqnarray}
	If the incoming state is strongly squeezed with $r\sim 1$, we have to restore the hyperbolic functions. As indicated above FIG.~\ref{FeynSum}, this represent the summation of an infinite amount of Feynman diagrams in the UV. 
	
	Notice that the terms proportional to $e^{2iz_\Lambda}$ is strongly cutoff dependent. A shift of $z_\Lambda\rightarrow z_\Lambda+\mathcal{O}(1)$ can change its value significantly. This strong dependence on the cutoff reflects the fact of energy conservation violation because of no incoming particles in the UV. And we regard it as due to the limitation of our model. In a simple quantum mechanics scattering problem, there exist results that are independent on the position of the potential barrier and those that are not. For instance, the transmittance and the reflectivity are independent of the position of the potential barrier while the transmitted wave function and the reflected wave function, however, are strongly dependent on the it. And they are all physical quantities (though some may not be directly observable). Similarly, the first term in (\ref{clock1}) and (\ref{clock2}) are weakly cutoff dependent while the second terms are strongly dependent.
	
	Clearly the clock signals originate from the four diagrams in FIG.~\ref{WQSFIArrows} that are untagged. They are suppressed by $\cosh r\sinh r\sim r$ and $\sinh^2 r\sim r^2$ instead of an exponential factor $\exp(-\pi \mu)$ which is usually too small for heavy fields with $m\gtrsim 10H$. Also notice that although the amplitudes for heavy massive fields are lifted in the squeezed limit, the equilateral-shape non-Gaussianity does not receive much uplift from the slightly non-BD initial condition because the analytical part comes not from quantum interferences. Therefore the signal-to-noise ratio is highly enhanced for massive fields with large masses, making them possible to be observed.
	
	\subsection{time-ordered part}
	In this section, we would like to evaluate the time ordered part contribution. Note that here the time ordered part is a bit different from the definitions in the original quasi-single field inflation paper because here we put the three-point interaction later than the two-point interaction in (\ref{rawclock})
	\begin{align}\nonumber
	\text{(time-ordered part)} = \Theta \left(\tau _1-\tau _2\right) & \times \left[\frac{\partial u_{k_1}\left(\tau _1\right){}^*}{\partial \tau _1} \frac{\partial u_{k_2}\left(\tau
		_1\right){}^*}{\partial \tau _1} \frac{\partial u_{k_3}\left(\tau _2\right){}^*}{\partial \tau _2} u_{k_1}(0) u_{k_2}(0) u_{k_3}(0) v_{k_3}\left(\tau _1\right) v_{k_3}\left(\tau
	_2\right){}^*\right.\\
	\nonumber& \left. -\frac{\partial u_{k_1}\left(\tau _1\right){}^*}{\partial \tau _1} \frac{\partial u_{k_2}\left(\tau _1\right){}^*}{\partial \tau _1} \frac{\partial
		u_{k_3}\left(\tau _2\right){}^*}{\partial \tau _2} u_{k_1}(0) u_{k_2}(0) u_{k_3}(0) v_{k_3}\left(\tau _2\right) v_{k_3}\left(\tau _1\right){}^*\right]+\text{c.c.}~.
	\end{align}
	Put it into integral, we have, for the first integral,
	\begin{align} \nonumber
	\mathcal T_1 &  = -    C_3 C_2   u_{k}(0)u_{k}(0)u_{k_3}(0)  \int_{\tau_\Lambda}^0 d\tau_1 \int_{\tau_\Lambda}^{\tau_1} d\tau_2   a^2 (\tau_1) u'^*_{k}(\tau_1) u'^*_{k}(\tau_1)  v_{k_3}(\tau_1) a^3(\tau_2) u'^*_{k_3}(\tau_2) v^*_{k_3}(\tau_2) +\text{c.c.}    \\ \nonumber 
	& =    C_3 C_2 \frac{ e^{-\pi\mu} H^3 \pi}{32k^2 k_3}  \int_{0}^{z_\Lambda/k_3} d x_1 \int_{x_1}^{z_\Lambda/k_3} dx_2   e^{-i k_3x_2 - 2 i k  {x_1}}  x_1^{3/2} x_2^{-1/2} H^{(1)}_{i\mu} ( k_3 {x_1}) H^{(2)}_{-i\mu} ( k_3 {x_2})  +\text{c.c.}    \\ \nonumber
	& =     C_3 C_2 \frac{ e^{-\pi\mu} H^3 \pi}{32k^2 k_3 }  \int_{0}^{z_\Lambda/k_3} d x_2 \int_{0}^{x_2} dx_1   e^{-i k_3x_2 - 2 i k  {x_1}}  x_1^{3/2} x_2^{-1/2} H^{(1)}_{i\mu} ( k_3 {x_1}) H^{(2)}_{-i\mu} ( k_3 {x_2}) +\text{c.c.}    \\ \nonumber
	& =   C_3 C_2 \frac{ e^{-\pi\mu} H^3 \pi}{32k^2 k_3}  \int_{0}^{z_\Lambda/k_3} d x_2 \int_{0}^{x_2} dx_1   e^{-i k_3x_2 - 2 i k  {x_1}}  x_1^{3/2} x_2^{-1/2} \bigg( \frac{2^{-i \mu } (\coth (\pi  \mu )+1) ( {k_3} {x_1})^{i \mu
	}}{\Gamma (i \mu +1)} \\ \nonumber
	& -\frac{i 2^{i \mu } \Gamma (i \mu ) ( {k_3}
		{x_1})^{-i \mu }}{\pi } \bigg) H^{(2)}_{-i\mu} ( k_3 {x_2})  +\text{c.c.}    \\ \nonumber
	& =   C_3 C_2 \frac{ e^{-\pi\mu} H^3 \pi}{32k^2 k_3}  \int_{0}^{z_\Lambda/k_3} d x_2 e^{-i k_3 x_2 } x_2^{-1/2} \bigg[  \frac{2^{-i \mu } (\coth (\pi  \mu )+1) ( {k_3} )^{i \mu
	}}{\Gamma (i \mu +1)} (2ik)^{-\frac{5}{2}-i\mu}   \gamma\bigg(\frac{5}{2}+i\mu,2ikx_2\bigg) \\
	& -\frac{i 2^{i \mu } \Gamma (i \mu ) ( {k_3}
		)^{-i \mu }}{\pi } (2ik)^{-\frac{5}{2}+i\mu}   \gamma\bigg(\frac{5}{2}-i\mu,2ikx_2\bigg)  \bigg] H^{(2)}_{-i\mu} ( k_3 {x_2}) +\text{c.c.}  ~.
	\end{align}
	The for the second equality, we redefine $x=-\tau$. We change the sequence of integration in the third equality and put the three-point interaction vertex to be integrated first. We do an IR expansion of the Hankel function in the fourth equality. 
	
	Making use the power series representation of the incomplete gamma function
	\begin{align}
	\gamma \bigg(\frac{5}{2}+i\mu,2ikx_2\bigg) = (2ikx_2)^{\frac{5}{2}+i\mu} \sum_{m=0}^{\infty} \frac{(-2ikx_2)^m}{m!(s+m)!}~.
	\end{align}
	We can see that the incomplete gamma function does not contribute to the clock signal. This is because the $k^{i\mu}$ factor from the series expansion of the incomplete gamma function cancels with the $k^{-i\mu}$ factor coming from the Hankel expansion. What is left are just analytical powers of $k$ and an integral only involving $k_3$. It cannot produce the clock signal of the form $(k_3/k)^{i\mu}=c^{i\mu}$.
	The same argument applies to the second integral as well.
	
	\section{The four-point function}\label{4ptsection}
	In addition to the primordial clock signals in the bispectrum, there exists analogous oscillatory behavior in the collapsed limit of the trispectrum. With the $\phi'^{2}\sigma$ interaction, the signals come from the diagram in which two inflatons exchange a massive field excitation. The non-analytic scalings of the ratio of comoving momentums can only arise from the IR expansion (\ref{IRexpansion}). As a result, we expand both interaction vertices in the limit of small $z_i=-k_I\tau_i,~i=1,2$. This is equivalent to saying that we assume the interaction mainly takes place late enough so that the massive field excitations become nonrelativistic and underdamped.
	
\begin{figure}[h!] 
		\centering 
		\includegraphics[width=15cm]{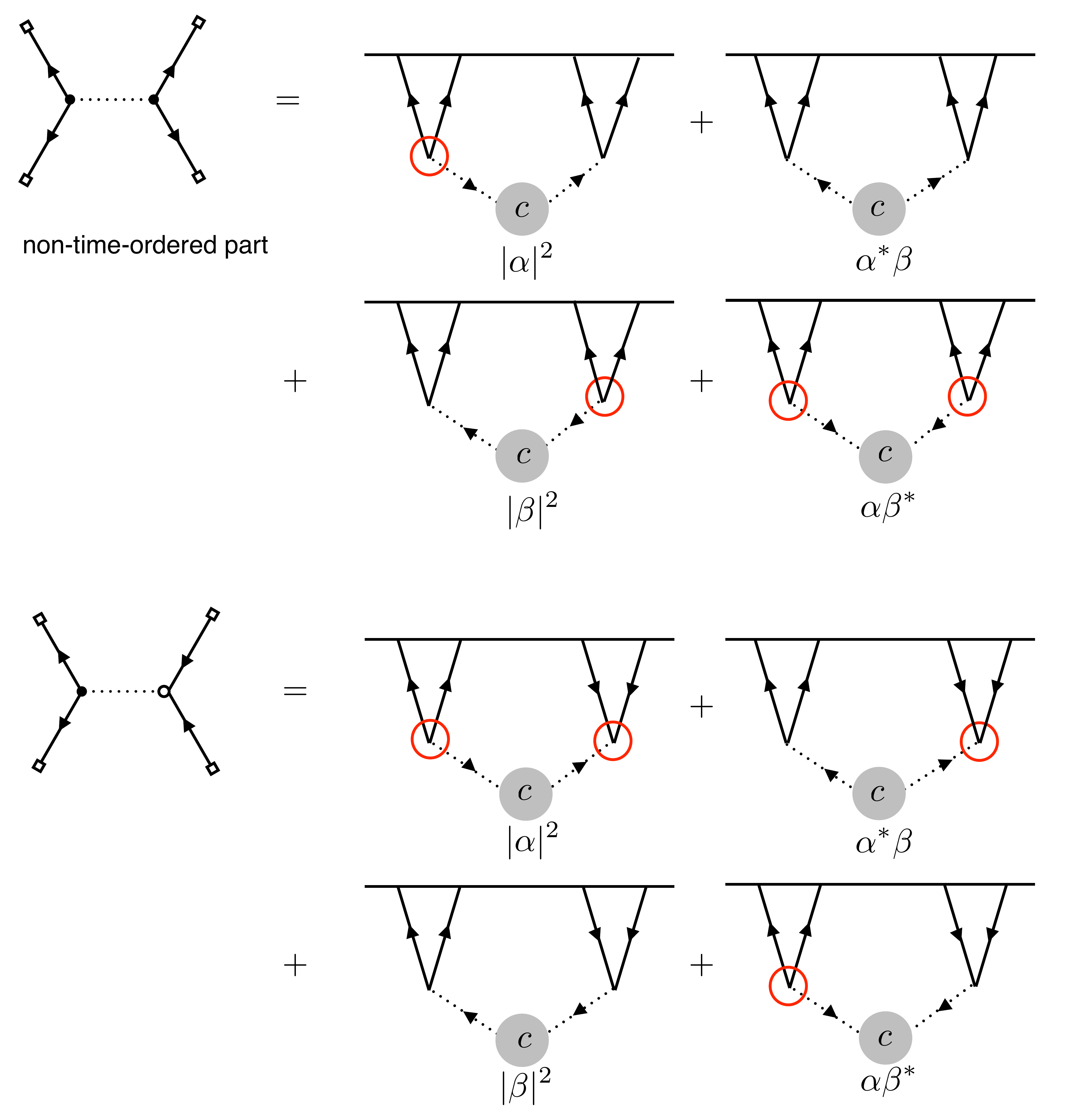}   
		\caption{Diagrammatic analysis of warm quasi-single field inflation four-point diagram. The two subdiagrams left untagged give unsuppressed clock signals that finally survived through the coarse-graining procedure.} \label{WQSFI4pt}
	\end{figure}

	By the same diagrammatic technique mentioned above, we can easily see that the clock signals only come from the $\alpha^*\beta\sim\mathcal{O}(r)$ diagram and the $|\beta|^2\sim \mathcal{O}(r^2)$ diagram indicated in FIG.~\ref{WQSFI4pt}
	
	\begin{eqnarray}\label{4ptalphabeta}
	&&\nonumber\langle\phi_{\mathbf{k}_1}\phi_{\mathbf{k}_2}\phi_{\mathbf{k}_3}\phi_{\mathbf{k}_4}\rangle'\equiv T_\phi(k_1,k_2,k_3,k_4,k_I)=\int \frac{d\tau_1}{(-H\tau_1)^2} \frac{d\tau_2}{(-H\tau_2)^2} C_3^2\\
	&&\times\left[(-1)^{a+b}\partial_{\tau_1}G_{a,1-a}(k_1,\tau_1,0)\partial_{\tau_1}G_{a,1-a}(k_2,\tau_1,0)D_{ab}(k_I,\tau_1,\tau_2)\times\partial_{\tau_2}G_{b,1-b}(k_3,\tau_2,0)\partial_{\tau_2}G_{b,1-b}(k_4,\tau_2,0)\right].~~~~~~~~
	\end{eqnarray}\label{4ptbetabeta}
	
	\begin{eqnarray}
	\text{(integrand)}&=&\alpha ^* \beta  \frac{\partial u_{k_1}\left(\tau _1\right)}{\partial \tau _1} \frac{\partial u_{k_2}\left(\tau _1\right)}{\partial \tau _1}
	\frac{\partial u_{k_3}\left(\tau _2\right)}{\partial \tau _2} \frac{\partial u_{k_4}\left(\tau _2\right)}{\partial \tau _2} u_{k_1}(0){}^* u_{k_2}(0){}^* u_{k_3}(0){}^* u_{k_4}(0){}^*
	v_{k_{\text{I}}}\left(\tau _1\right){}^* v_{k_{\text{I}}}\left(\tau _2\right){}^*~~~~~~~~~\\
	&&-|\beta| ^2 \frac{\partial u_{k_3}\left(\tau _2\right)}{\partial
		\tau _2} \frac{\partial u_{k_4}\left(\tau _2\right)}{\partial \tau _2} \frac{\partial u_{k_1}\left(\tau _1\right){}^*}{\partial \tau _1} \frac{\partial u_{k_2}\left(\tau
		_1\right){}^*}{\partial \tau _1} u_{k_1}(0) u_{k_2}(0) u_{k_3}(0){}^* u_{k_4}(0){}^* v_{k_{\text{I}}}\left(\tau _1\right) v_{k_{\text{I}}}\left(\tau _2\right){}^*\\
	\nonumber&&\Theta \left(\tau _1-\tau _2\right) \\
	\nonumber&&\times\left[\frac{\partial u_{k_1}\left(\tau _1\right){}^*}{\partial \tau_1}\frac{\partial u_{k_2}\left(\tau _1\right){}^*}{\partial \tau _1} \frac{\partial u_{k_3}\left(\tau _2\right){}^*}{\partial \tau _2} \frac{\partial u_{k_4}\left(\tau
		_2\right){}^*}{\partial \tau _2} u_{k_1}(0) u_{k_2}(0) u_{k_3}(0) u_{k_4}(0) v_{k_{\text{I}}}\left(\tau _1\right) v_{k_{\text{I}}}\left(\tau _2\right){}^*\right.\\
	&&\left.-\frac{\partial u_{k_1}\left(\tau _1\right){}^*}{\partial \tau _1} \frac{\partial u_{k_2}\left(\tau _1\right){}^*}{\partial \tau _1}
	\frac{\partial u_{k_3}\left(\tau _2\right){}^*}{\partial \tau _2} \frac{\partial u_{k_4}\left(\tau _2\right){}^*}{\partial \tau _2} u_{k_1}(0) u_{k_2}(0) u_{k_3}(0) u_{k_4}(0)
	v_{k_{\text{I}}}\left(\tau _2\right) v_{k_{\text{I}}}\left(\tau _1\right){}^*\right]\\
	\nonumber&&+\text{c.c.}~.
	\end{eqnarray}
	The final trispectrum is
	\begin{equation}
	T_\phi(k_1,k_2,k_3,k_4,k_I)=C_3^2\int_{-z_\Lambda/k_I}^{0}\frac{d\tau_1}{(-H\tau_1)^2}\int_{-z_\Lambda/k_I}^{0}\frac{d\tau_2}{(-H\tau_2)^2}\times\left[\text{(integrand)}\right]_{k_I/k_i\rightarrow 0}~.
	\end{equation}
	We follow the methodology above and evaluate the integrand separately.
	\subsection{non-time-ordered part}
	Direct integration gives
	\begin{eqnarray}
	T_\phi(k_1,k_2,k_3,k_4,k_I)&=&\alpha ^* \beta T_\phi(k_1,k_2,k_3,k_4,k_I)^{(1)}+|\beta|^2T_\phi(k_1,k_2,k_3,k_4,k_I)^{(2)}~.
	\end{eqnarray}
	under the assumption of the hierarchy $1\ll\mu\ll z_\Lambda$. With the Bogolyubov coefficients obtained in (\ref{BogolyubovCoeff}), the leading order clock signal and the next order is
	\begin{eqnarray}
	T_\phi(k_1,k_2,k_3,k_4,k_I)^{(1)}&=&-C_3^2\frac{i \pi  H^6 \mu ^3   }{16 k_1 k_2 k_3 k_4 \left(k_1+k_2\right){}^{5/2} \left(k_3+k_4\right){}^{5/2}}\left(\frac{k_{I}/2}{k_1+k_2}\right)^{-i \mu } \left(\frac{k_{I}/2}{k_3+k_4}\right)^{-i \mu }\\
	\nonumber&&+\text{c.c.}\\
	T_\phi(k_1,k_2,k_3,k_4,k_I)^{(2)}&=&-C_3^2 \frac{ \pi H^6 \mu ^3  }{16 k_1 k_2 k_3 k_4 \left(k_1+k_2\right){}^{5/2} \left(k_3+k_4\right){}^{5/2}}\left(\frac{k_3+k_4}{k_1+k_2}\right)^{-i \mu }\\
	\nonumber&&+\text{c.c.}~.
	\end{eqnarray}
	Taking into account higher-order terms, we get
	\begin{eqnarray}
	T_\phi(k_1,k_2,k_3,k_4,k_I)&=&e^{2iz_\Lambda}\cosh r\sinh r T_\phi(k_1,k_2,k_3,k_4,k_I)^{(1)}+\sinh^2 r T_\phi(k_1,k_2,k_3,k_4,k_I)^{(2)}~.
	\end{eqnarray}
	
	As a result, the trispectrum in warm quasi-single field inflation scenario shows two types of clock signals that are not suppressed by the large-mass exponential $e^{-\pi m/H}$. Moreover, at the first order, only clock signal of the type $(k_I/(k_1+k_2))^{\pm i\mu}(k_I/(k_3+k_4))^{\pm i\mu}$ shows up while at the second order, only clock signal of the type $((k_1+k_2)/(k_3+k_4))^{\pm i\mu}$ is present.
	
	\subsection{time-ordered part}
	The time-ordered part is evaluated at each vertex using the large $\mu$ approximation.
	\begin{eqnarray}
	T^\text{time-ordered}&=&C_3^2\int_{-z_\Lambda/k_I}^{0}\frac{d\tau_2}{(-H\tau_2)^2}\left[\int_{-z_\Lambda/k_I}^{\tau_2}\frac{d\tau_1}{(-H\tau_1)^2}\times\left[\text{(time-ordered part)}\right]_{k_I/k_i\rightarrow 0}\right]_{\mu\rightarrow\infty}\\
	&=&\text{(analytical)}+\text{(non-analytical)}~.
	\end{eqnarray}
	The final expression is a mixture of analytical terms and non-analytical terms which in the usual case would lead to the clock signals. However, these terms contain a factor
	\begin{equation}
	\exp\left(\pm i\frac{k_3+k_4}{k_I}z_\Lambda\right)
	\end{equation}
	that oscillates excessively fast and are effectively invisible. Thus this part is not as observable as the non-time-ordered part.
	
	\section{Conclusion}
	We study a parameter regime of warm inflation, where the clock signal produced by the massive field can evade the large mass Boltzmann suppression. A considerable amount of particles are generated in the background thermal bath. This offers us an opportunity to probe the UV physics of the early universe if inflation indeed happened with a temperature higher than Hubble.
	
	The UV physics inside the thermal bath is described by the flat space thermal field theory. The thermal bath gives rise to the mass correction of the primordial fields. The effect of the thermal bath is to squeeze the BD vacuum, and thus provides a squeezed state from the UV furnace. It can also be described by an effective vertex in the in-in formalism. The usual mass Boltzmann factor becomes a suppression on the energy of radiation in this high temperature background. After the modes evolve into the IR region, they exit the thermal equilibrium and the physics there can be described by the usual in-in formalism.   
	
	The change of mass induces particle production. This process has the same mathematical structure as a scattering problem of quantum mechanics. And the transmission rate corresponds to the particle production rate. We demonstrated the physics with this intuitive picture.
	
	Later we calculated the bispectrum and trispectrum in our model and showed explicitly that the high temperature effect can enhance the clock signal by a significant amount. We also used a simple estimator of the clock signal and show that the enhanced clock signal is indeed due to the particle production at the UV furnace and not an artifact from connecting the UV and IR theories.  
	
	We have not yet considered the effect from the particles which originally existed in the thermal bath. The formalism we developed here cannot deal with the signal produced by this type of particles. The signals produced by them have completely different origin with the signals we considered here, so they cannot cancel the signal that we have calculated. We hope to study these contributions in the future.
	
	Our result offers a possibility of probing highly UV physics such as string theory. String theory usually produces massive higher spin fields of masses much larger than the Hubble scale. In most part of the parameter space, the non-analytic signature is likely to be small and hard to observe. Our result shows that the signal can be much larger than we previously thought, as long as the high-temperature warm inflation background is established.

	\section*{Acknowledgments} 
	We would like to thank Hayden Lee, Shi Pi, Misao Sasaki and Jiro Soda for helpful discussions and comments. This work was supported by ECS Grant 26300316 and GRF Grant 16301917 from the Research Grants Council of Hong Kong. XT is supported by the Qian San-Qiang Class in the University of Science and Technology of China. SZ is supported by the the Hong Kong PhD Fellowship Scheme (HKPFS) issued by the Research Grants Council (RGC) of Hong Kong.

	\appendix
	\section{mass corrections}\label{AppendixA}
	There are many interactions leading to the thermal correction of the mass of the isocurvaton field $\sigma$. Here we focus on three primary ones in consideration of operator dimensions and the effective theory setup (\ref{EFTsetup}). Namely, they are the triple and quartic self-interactions of the isocurvaton and the interaction with the inflaton $\phi$.
	\begin{equation}
	\Delta \mathcal{L}=\sqrt{-g}\left(-\frac{\varrho}{3!}\sigma^3-\frac{\lambda}{4!}\sigma^4-\frac{1}{2\Lambda_0} (\partial \phi)^2 \sigma\right)~.
	\end{equation}
	Since the thermal processes happen in the UV furnace, the characteristic time and length scale over which the processes take place is significantly smaller than that of the curvature of spacetime. In other words, the metric does not change much during the whole process. Thus we can ignore the overall $\sqrt{-g}$ and choose the scale factor $a=1$ so that the space coordinate is approximately physical during the process we are considering. In this way, it is reasonable to directly apply finite temperature field theory in flat spacetime. We use real-time formalism below and dress the propagators with a thermal term.
	\begin{enumerate}[$\bullet$]
		\item Triple self-interaction
		
		The triple interaction is present if no reflection symmetry $\sigma\leftrightarrow-\sigma$ exists. The one loop diagram reads
		\begin{eqnarray}
		\nonumber-i\mathcal{M}_3^2(p^2)&=&-\frac{\varrho^2}{2}\int\frac{d^4 k}{(2\pi)^4}\left(\frac{i}{-k^2-m^2}+2\pi n_B(E_k)\delta(-k^2-m^2)\right)\\
		&&~~~~~~~~~~~~~~~\times\left(\frac{i}{-(p+k)^2-m^2}+2\pi n_B(E_{p+k})\delta\left(-(p+k)^2-m^2\right)\right)~.
		\end{eqnarray}
		The mass correction is given by the cross term. In general, the mass correction is dependent on the three-momentum of the isocurvaton, which slightly modify the dispersion relationship. We focus on the mass correction at rest:
		\begin{eqnarray}
		\nonumber\Delta m_3^2(T)&=&\varrho^2\int\frac{d^4 k}{(2\pi)^4}\frac{2\pi}{e^{\beta E_{k}}-1}\frac{\delta(-k^2-m^2)}{-(p+k)^2-m^2}\\
		\nonumber&&\xrightarrow{p=(m,\mathbf{0})}-\frac{\varrho^2}{2\pi^2}\int_0^\infty d|\mathbf{k}|\dfrac{|\mathbf{k}|^2/\sqrt{|\mathbf{k}|^2+m^2}}{4|\mathbf{k}|^2+3m^2}\frac{1}{e^{\beta\sqrt{|\mathbf{k}|^2+m^2}}-1}\\
		&\simeq&-\frac{\varrho^2 T}{30\pi m}~.
		\end{eqnarray}
		
		\item Quartic self-interaction
		
		The quartic self-interaction induces a mass correction that is independent of the three-momentum of the isocurvaton.
		\begin{eqnarray}
		-i\mathcal{M}_4^2(p^2)&=&-\frac{i\lambda}{2}\int\frac{d^4 k}{(2\pi)^4}\left(\frac{i}{-k^2-m^2}+2\pi n_B(E_k)\delta(-k^2-m^2)\right)~.
		\end{eqnarray}
		\begin{eqnarray}
		\nonumber\Rightarrow \Delta m_4^2(T)&=&\frac{\lambda}{(2\pi)^2}\int d|\mathbf{k}|\frac{|\mathbf{k}|^2}{\sqrt{|\mathbf{k}|^2+m^2}}\frac{1}{e^{\beta\sqrt{|\mathbf{k}|^2+m^2}}-1}\\
		&=&\frac{\lambda T^2}{24}\times\left(1+\mathcal{O}(\frac{m}{T})\right)\simeq\frac{\lambda T^2}{24}~.
		\end{eqnarray}
		
		\item Interaction with the inflaton
		The interaction with the massless inflaton is dimension-five and the coupling constant is the inverse of a energy scale $\Lambda_0$. The mass correction comes from an inflaton loop at the leading order.
		\begin{eqnarray}
		\nonumber-i\mathcal{M}_I^2(p^2)&=&-\frac{1}{2\Lambda_0^2}\int\frac{d^4 k}{(2\pi)^4}\left[-k\cdot(p+k)\right]^2\left(\frac{i}{-k^2+i\epsilon}+2\pi n_B(E_k)\delta(-k^2)\right)\\
		&&~~~~~~~~~~~~~~~\times\left(\frac{i}{-(p+k)^2+i\epsilon}+2\pi n_B(E_{p+k})\delta\left(-(p+k)^2\right)\right)~.
		\end{eqnarray}
		Again the thermal correction to the mass coming from the cross term is dependent on the three-momentum. Focusing on the rest mass, we obtain
		\begin{eqnarray}
		\nonumber\Rightarrow \Delta m_I^2(T)&=&\frac{1}{\Lambda_0^2}\int \frac{d^4 k}{(2\pi)^4}\left[-k\cdot(p+k)\right]^2\frac{\delta(-k^2)}{-(p+k)^2+i\epsilon}\frac{2\pi}{e^{\beta E_k}-1}\\
		\nonumber&&\xrightarrow{p=(m,\mathbf{0})}-\frac{m^2}{2\pi^2\Lambda_0^2}\int_0^\infty d|\mathbf{k}|\dfrac{|\mathbf{k}|^3/(4|\mathbf{k}|^2-m^2-i\epsilon)}{e^{\beta|\mathbf{k}|}-1}\\
		&&\simeq-\frac{m^2T^2}{48\Lambda_0^2}~.
		\end{eqnarray}
		where the pole in the integrand originates from the on-shell production of the isocurvaton $\sigma$. And the principal value is assumed during integration.
	\end{enumerate} 
	
	\section{A check on the cause of unsuppressed clock signals}\label{alltimesection}
	Here in this part of the appendix we show the necessity for the two-point interaction to only happen in the extreme UV but not throughout the history. A BD initial condition is assumed and we manually add a two-point interaction of the massive field that can either last for a short time in the UV or throughout inflation.
	
	To identify the existence of clock signals, we check the norm of equal-time two-point function of the massive field $|\langle\sigma_\mathbf{k}(\tau)\sigma_\mathbf{-k}(\tau)\rangle|$ and see if it oscillates with time. If it does, then there should be unsuppressed clock signals, otherwise not.
	
	The equal-time two-point function is to the lowest order the diagram below.
	\begin{figure}[htbp] 
		\centering 
		\includegraphics[width=5cm]{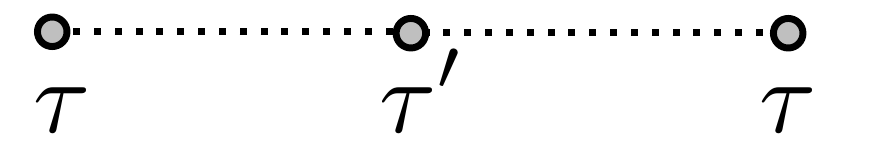}   
		\caption{Equal-time two-point function.} 
	\end{figure}
	
	Set $\tau=\tau_1=\tau_2$ and denote the interaction time $\tau'$. It is easy to check the four cases give the same answer and are thus indistinguishable,
	\begin{eqnarray}
	D_{-a}D_{a+}=D_{+a}D_{a-}=\frac{D_{+a}D_{a+}+D_{-a}D_{a-}}{2}~.
	\end{eqnarray}
	Then the equal-time two-point function is essentially
	\begin{eqnarray}
	\langle\sigma_\mathbf{k}(\tau)\sigma_\mathbf{-k}(\tau)\rangle\propto D_{-a}D_{a+}\propto\int \frac{d\tau'}{(-H\tau')^4} \theta(\tau-\tau')\epsilon(\tau')\left[v_k(\tau')^2 v_k(\tau)^{*2}-v_k(\tau')^{*2} v_k(\tau)^2\right]~.
	\end{eqnarray}
	Plug in the explicit expression for the mode function and omit all the irrelevant coefficients, and we arrive at the crucial result,
	\begin{eqnarray}\label{eq2pt}
	2 i {\rm Im} \left(z^3 H_{-i \mu}^{(2)}(z){}^2 \int_{z}^{\infty } \frac{dz'}{z'} \, H_{i
		\mu}^{(1)}(z'){}^2\epsilon(z')\right)~.
	\end{eqnarray}
	The indefinite integral turns out to be
	\begin{eqnarray}
	\int \frac{dz'}{z'} \, H_{i
		\mu}^{(1)}(z'){}^2
	&=&\frac{i 2^{-1-2 i \mu} z'^{2 i \mu} (\coth (\pi  \mu)+1)^2 \, _2F_3\left(i
		\mu+\frac{1}{2},i \mu;i \mu+1,i \mu+1,2 i \mu+1;-z'^2\right)}{\mu^3 \Gamma (i
		\mu)^2}-\\
	\nonumber&&\frac{i 2^{-1+2 i \mu} z'^{-2 i \mu} \text{csch}^2(\pi  \mu) \,
		_2F_3\left(\frac{1}{2}-i \mu,-i \mu;1-i \mu,1-i \mu,1-2 i \mu;-z'^2\right)}{\mu^3
		\Gamma (-i \mu)^2}-\\
	\nonumber&&\frac{(\coth (\pi  \mu)+1) \left(-z'^2 \,
		_3F_4\left(1,1,\frac{3}{2};2,2,2-i \mu,i \mu+2;-z'^2\right)+4 \mu^2 \log (z')+4
		\log (z')\right)}{2 \pi  \mu (\mu-i) (\mu+i)}~.
	\end{eqnarray}
	There are two scenarios:
	\begin{enumerate}
		\item The effective two-point vertex lasts the whole time, $\epsilon(z')=1$.
		\item The effective two-point vertex only exists above $z_\Lambda$, $\epsilon(z')=\theta(z'-z_\Lambda)$.
	\end{enumerate}
	
	Instead of explicitly writing the messy and unenlightening results for each case, we plot them on the same figure and the difference is very clear. Case 1 is without oscillation while case 2 is. So we conclude that if the interaction only happens in the extreme UV regime, the clock signal will not be suppressed by a low Hubble temperature.
	\begin{figure}[htbp] 
		\centering 
		\includegraphics[width=18cm]{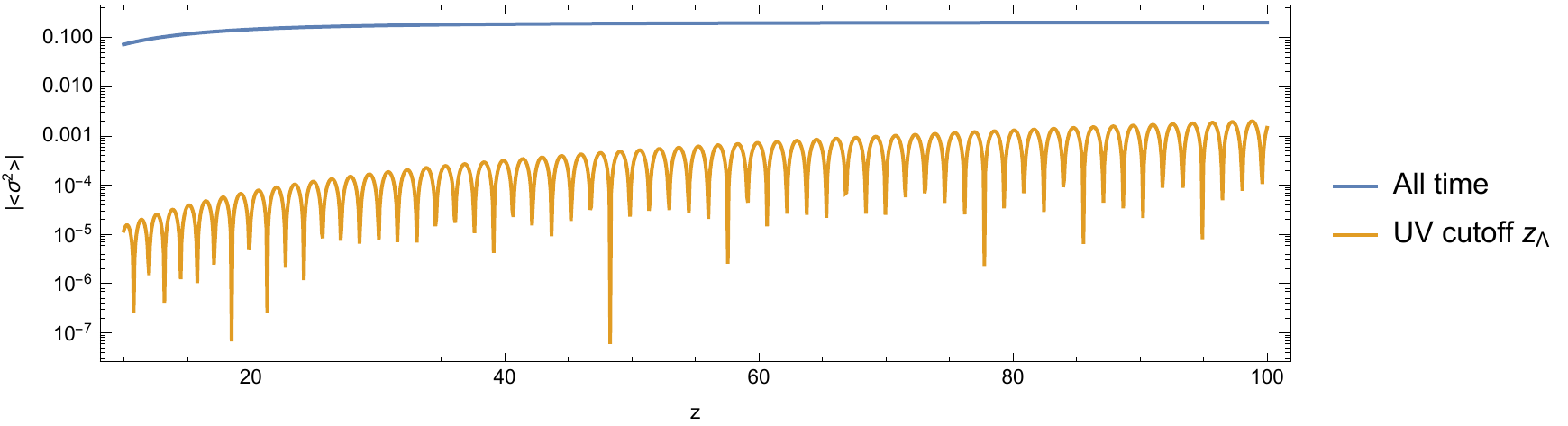}   
		\caption{Comparison between the two cases. Here we took $z_\Lambda=1000$ and $m=10H$.} 
	\end{figure}
	
	The cause for the disappearance of the clock can be argued as follows. In (\ref{eq2pt}) we see that $H_{-i \mu}^{(2)}(z){}^2\sim e^{-2iz}$ is oscillatory. If there are no restriction for the integrand (i.e. $\epsilon(z')=1$), the integral goes like
	\begin{equation}
	\int_{z}^{\infty } \frac{dz'}{z'} \, H_{i\mu}^{(1)}(z'){}^2\sim e^{2iz},
	\end{equation}
	resulting in the cancellation between the two oscillations. But with a cutoff $z_\Lambda$, the integral is essentially a constant. As a result, the oscillation in $z$ comes solely from $H_{-i \mu}^{(2)}(z){}^2\sim e^{-2iz}$.
	
	\section{In in formalism and effective vertex method}\label{AppendixC}
	In this section, we would like to introduce a new method to study the initial conditions, which is the effective vertex method. We can see that the changing of mass at $z_\Lambda$ is equivalent to imposing a non-BD initial condition at $z_\Lambda$.
	
	The interaction is composed of three parts,
	\begin{align}
	H_t = \tilde H_I + H_2 + H_3~,
	\end{align}
	where $\tilde H_I$ is already defined in \eqref{interactionhamiltonian}, $H_2$ and $H_3$ are two-point and three-point interaction vertexes of the inflaton and massive field, respectively, 
	\begin{align}
	H_2 = - a^3 C_2 \int d^3 x \sigma \phi', \quad H_3 = - a^2 C_3 \int d^3 x \sigma \phi' \phi'~.
	\end{align}
	From in-in formalism, we can calculate the late time observer $Q$ ($Q=\phi\phi\phi$ for three-point functions and $Q=\phi\phi\phi\phi$ for four-point functions) order by order as
	\begin{align} \nonumber
	\langle Q \rangle  = & -2 {\rm Im} \langle 0 | Q \int_{-\infty}^0 d\tau_1 \int_{-\infty}^{\tau_1} d\tau_2 \int_{-\infty}^{\tau_2} d\tau_3 H_{t}(\tau_1) H_{t}(\tau_2) H_{t} (\tau_3) | 0 \rangle  \\
	&  +  2 {\rm Im} \langle 0 | \int_{-\infty}^0 d \tilde \tau_1 H_{t}(\tilde\tau_1) Q \int_{-\infty}^0 d \tau_{1} \int_{-\infty}^{\tau_1} d\tau_2 H_{t}(\tau_1 ) H_{t}(\tau_2) | 0 \rangle   \\\nonumber
	& +   \langle 0 | \int_{-\infty}^0 d\tilde \tau_1 \int_{-\infty}^{\tilde \tau_1} d\tilde \tau_2 H_{t}(\tilde \tau_2) H_{t}(\tilde \tau_1)  Q \int_{-\infty}^0 d\tau_1 \int_{-\infty}^{\tau_1} d\tau_2  H_{t}(\tau_1) H_{t}(\tau_2)  | 0 \rangle  \\ \nonumber
	&  -  2 {\rm Re} \langle 0 | \int_{-\infty}^0 d \tilde \tau_1 H_{t}(\tilde\tau_1) Q \int_{-\infty}^0 d \tau_{1} \int_{-\infty}^{\tau_1} d\tau_2 \int_{-\infty}^{\tau_2} d\tau_3 H_{t}(\tau_1 ) H_{t}(\tau_2) H_{t}(\tau_3) | 0 \rangle \\
	& + 2 {\rm Re} \langle 0 | Q \int_{-\infty}^0 d\tau_1 \int_{-\infty}^{\tau_1} d\tau_2 \int_{-\infty}^{\tau_2} d\tau_3 \int_{-\infty}^{\tau_3} d\tau_4 H_{t}(\tau_1) H_{t}(\tau_2) H_{t} (\tau_3) H_{t} (\tau_4) | 0 \rangle   ~.
	\end{align}
	The first two lines are expansion of the Hamiltonian to the third order. The last three lines are expansion of the Hamiltonian to the forth order.
	
	\subsection{3pt} 
	In this subsection, we consider the in in formalism to compute the three-point correlation function. 
	\subsubsection{$\mathcal{O}(r)$ Order}
	\begin{figure}[h!] 
		\centering 
		\includegraphics[width=5cm]{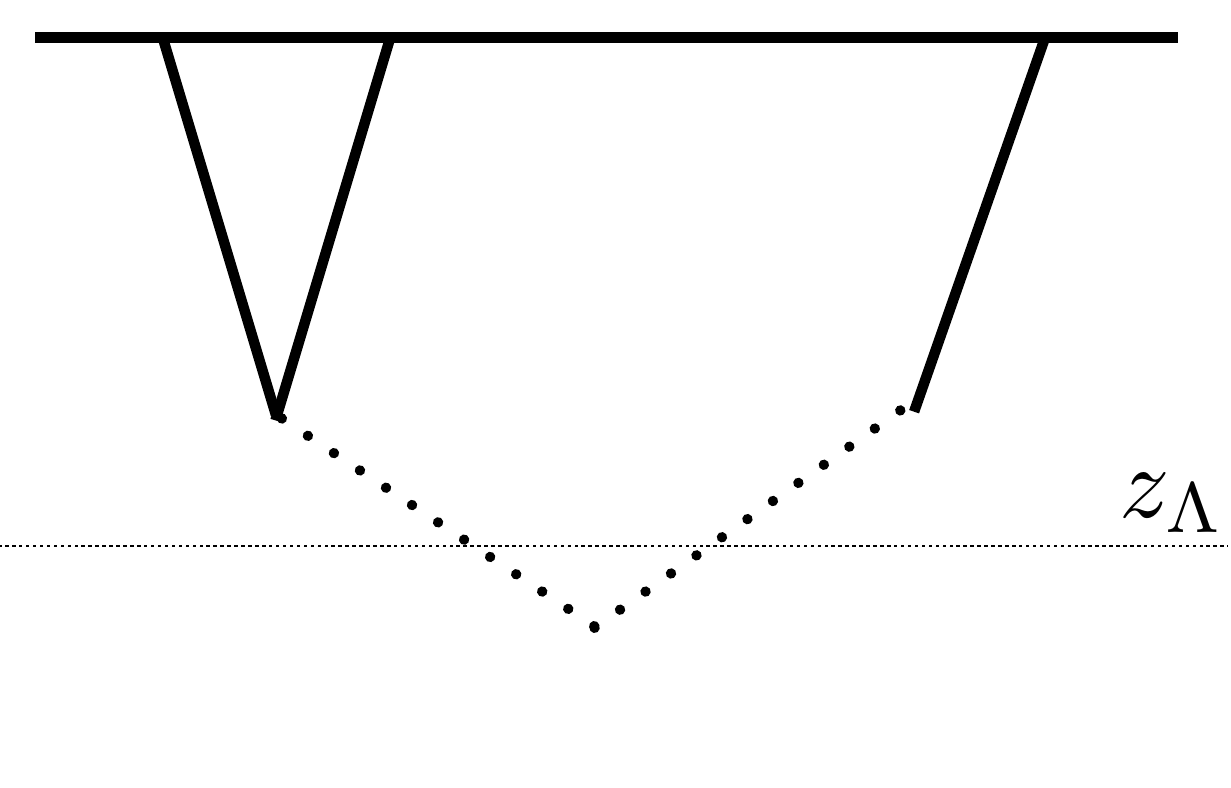}  
		\caption{Feynman diagram for the three-point correlation function at $r$ order. The upper horizontal line is the $\tau=0$ time slice, whereas the lower dashed horizontal line is the $-k\tau=z_\Lambda$ cutoff time below which the effective mass square of the $\sigma$ field is $m^2+\Delta m^2$ and above which the effective mass of the $\sigma$ field is $m^2$. The other three solid lines represent the inflaton field $\phi$ and the dashed lines corresponds to the massive field $\sigma$. } \label{ininformalism3ptr1}  
	\end{figure}
	The terms that actually contributes are
	\begin{align} \nonumber
	\langle \phi\phi\phi \rangle  = & -2 {\rm Im} \langle 0 | \phi\phi\phi \int_{\tau_\Lambda}^0 d\tau_1 \int_{\tau_\Lambda}^{\tau_1} d\tau_2 \int_{-\infty}^{\tau_\Lambda} d\tau_3 H_{2}(\tau_1) H_{3}(\tau_2) \tilde H_{I} (\tau_3) | 0 \rangle  \\\nonumber
	& -2 {\rm Im} \langle 0 | \phi\phi\phi \int_{\tau_\Lambda}^0 d\tau_1 \int_{\tau_\Lambda}^{\tau_1} d\tau_2 \int_{-\infty}^{\tau_\Lambda} d\tau_3 H_{3}(\tau_1) H_{2}(\tau_2) \tilde H_{I} (\tau_3) | 0 \rangle  \\\nonumber
	&  +  2 {\rm Im} \langle 0 | \int_{\tau_\Lambda}^0 d \tilde \tau_1 H_{2}(\tilde\tau_1) \phi\phi\phi \int_{\tau_\Lambda}^0 d \tau_{1} \int_{-\infty}^{\tau_\Lambda} d\tau_2 H_{3}(\tau_1 ) \tilde H_{I}(\tau_2) | 0 \rangle \\\nonumber
	&  +  2 {\rm Im} \langle 0 | \int_{\tau_\Lambda}^0 d \tilde \tau_1 H_{3}(\tilde\tau_1) \phi\phi\phi \int_{\tau_\Lambda}^0 d \tau_{1} \int_{-\infty}^{\tau_\Lambda} d\tau_2 H_{2}(\tau_1 ) \tilde H_{I}(\tau_2) | 0 \rangle \\\nonumber
	&  +  2 {\rm Im} \langle 0 | \int_{-\infty}^{\tau_\Lambda} d \tilde \tau_1 \tilde H_{I}(\tilde\tau_1) \phi\phi\phi \int_{\tau_\Lambda}^0 d \tau_{1} \int_{\tau_\Lambda}^{\tau_1} d\tau_2 H_{2}(\tau_1 ) H_{3}(\tau_2) | 0 \rangle\\
	&  +  2 {\rm Im} \langle 0 | \int_{-\infty}^{\tau_\Lambda} d \tilde \tau_1 \tilde H_{I}(\tilde\tau_1) \phi\phi\phi \int_{\tau_\Lambda}^0 d \tau_{1} \int_{\tau_\Lambda}^{\tau_1} d\tau_2 H_{3}(\tau_1 ) H_{2}(\tau_2) | 0 \rangle  ~.
	\end{align} 
	We need to first integrate out the $\tilde H_I$, the third and forth line evaluates to the form the same as the first line of \eqref{3ptalphabeta}, the first, second, fifth and sixth line evaluates to the form the same as the second line of \eqref{3ptalphabeta}.
	
	\subsubsection{$\mathcal{O}(r^2)$ Order}
	\begin{figure}[h!] 
		\centering 
		\includegraphics[width=5cm]{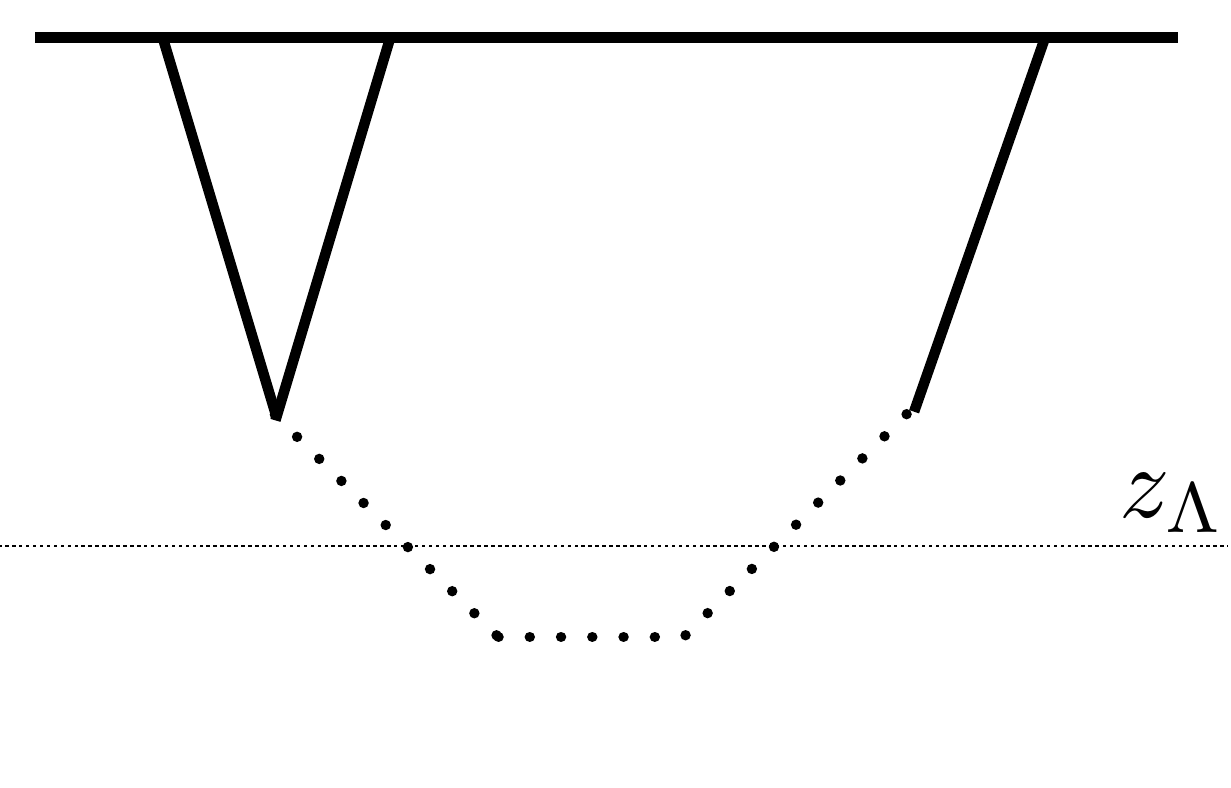}   
		\caption{Feynman diagram for the three-point correlation function at $r^2$ order. The upper horizontal line is the $\tau=0$ time slice, whereas the lower dashed horizontal line is the $k\tau=z_\Lambda$ cutoff time below which the effective mass square of the $\sigma$ field is $m^2+\Delta m^2$ and above which the effective mass of the $\sigma$ field is $m^2$. The other three solid lines represent the inflaton field $\phi$ and the dashed lines corresponds to the massive field $\sigma$. } \label{ininformalism3ptr2} 
	\end{figure}
	\begin{align} \nonumber
	\langle \phi\phi\phi \rangle  & =   \langle 0 | \int_{\tau_\Lambda}^0 d\tilde \tau_1 \int_{\tau_\Lambda}^{\tilde \tau_1} d\tilde \tau_2 H_{2}(\tilde \tau_2) H_{3}(\tilde \tau_1)  \phi\phi\phi \int_{-\infty}^{\tau_\Lambda} d\tau_1 \int_{-\infty}^{\tau_1} d\tau_2  \tilde H_{I} (\tau_1) \tilde H_{I}(\tau_2)  | 0 \rangle  \\ \nonumber
	& +   \langle 0 | \int_{\tau_\Lambda}^0 d\tilde \tau_1 \int_{\tau_\Lambda}^{\tilde \tau_1} d\tilde \tau_2 H_{3}(\tilde \tau_2) H_{2}(\tilde \tau_1)  \phi\phi\phi \int_{-\infty}^{\tau_\Lambda} d\tau_1 \int_{-\infty}^{\tau_1} d\tau_2  \tilde H_{I} (\tau_1)  \tilde H_{I} (\tau_2)  | 0 \rangle  \\ \nonumber
	& +   \langle 0 | \int_{\tau_\Lambda}^0 d\tilde \tau_1 \int_{-\infty}^{\tau_\Lambda} d\tilde \tau_2 H_{I}(\tilde \tau_2) \tilde H_{2} (\tilde \tau_1)  \phi\phi\phi \int_{\tau_\Lambda}^0 d\tau_1 \int_{-\infty}^{\tau_\Lambda} d\tau_2  H_{3} (\tau_1) \tilde H_{I} (\tau_2)  | 0 \rangle  \\ \nonumber
	& +   \langle 0 | \int_{\tau_\Lambda}^0 d\tilde \tau_1 \int_{-\infty}^{\tau_\Lambda} d\tilde \tau_2 H_{I}(\tilde \tau_2) \tilde H_{3} (\tilde \tau_1)  \phi\phi\phi \int_{\tau_\Lambda}^0 d\tau_1 \int_{-\infty}^{\tau_\Lambda} d\tau_2  H_{2}(\tau_1) \tilde H_{I} (\tau_2)  | 0 \rangle  \\ \nonumber
	& +   \langle 0 | \int_{-\infty}^{\tau_\Lambda} d\tilde \tau_1 \int_{-\infty}^{\tilde \tau_1} d\tilde \tau_2 \tilde H_{I} (\tilde \tau_2) \tilde H_{I} (\tilde \tau_1)  \phi\phi\phi \int_{\tau_\Lambda}^0 d\tau_1 \int_{\tau_\Lambda}^{\tau_1} d\tau_2  H_{2}(\tau_1) H_{3}(\tau_2)  | 0 \rangle  \\ \nonumber
	& +   \langle 0 | \int_{-\infty}^{\tau_\Lambda} d\tilde \tau_1 \int_{-\infty}^{\tilde \tau_1} d\tilde \tau_2 \tilde H_{I} (\tilde \tau_2) \tilde H_{I} (\tilde \tau_1)  \phi\phi\phi \int_{\tau_\Lambda}^0 d\tau_1 \int_{-\infty}^{\tau_1} d\tau_2  H_{3}(\tau_1) H_{2}(\tau_2)  | 0 \rangle  \\ \nonumber
	&  -  2 {\rm Re} \langle 0 | \int_{\tau_\Lambda}^0 d \tilde \tau_1 H_{2}(\tilde\tau_1) \phi\phi\phi \int_{\tau_\Lambda}^0 d \tau_{1} \int_{-\infty}^{\tau_\Lambda} d\tau_2 \int_{-\infty}^{\tau_2} d\tau_3 H_{3}(\tau_1 ) \tilde H_{I} (\tau_2) \tilde H_{I} (\tau_3) | 0 \rangle \\ \nonumber
	&  -  2 {\rm Re} \langle 0 | \int_{\tau_\Lambda}^0 d \tilde \tau_1 H_{3}(\tilde\tau_1) \phi\phi\phi \int_{\tau_\Lambda}^0 d \tau_{1} \int_{-\infty}^{\tau_\Lambda} d\tau_2 \int_{-\infty}^{\tau_2} d\tau_3 H_{2}(\tau_1 ) \tilde H_{I} (\tau_2) \tilde H_{I} (\tau_3) | 0 \rangle \\ \nonumber
	&  -  2 {\rm Re} \langle 0 | \int_{-\infty}^{\tau_\Lambda} d \tilde \tau_1 \tilde H_{I}(\tilde\tau_1) \phi\phi\phi \int_{\tau_\Lambda}^0 d \tau_{1} \int_{\tau_\Lambda}^{\tau_1} d\tau_2 \int_{-\infty}^{\tau_\Lambda} d\tau_3 H_{2}(\tau_1 ) H_{3}(\tau_2) \tilde H_{I}(\tau_3) | 0 \rangle \\ \nonumber
	&  -  2 {\rm Re} \langle 0 | \int_{-\infty}^{\tau_\Lambda} d \tilde \tau_1 \tilde H_{I} (\tilde\tau_1) \phi\phi\phi \int_{\tau_\Lambda}^0 d \tau_{1} \int_{\tau_\Lambda}^{\tau_1} d\tau_2 \int_{-\infty}^{\tau_\Lambda} d\tau_3 H_{3}(\tau_1 ) H_{2}(\tau_2) \tilde H_{I}(\tau_3) | 0 \rangle \\ \nonumber
	& + 2 {\rm Re} \langle 0 | \phi\phi\phi \int_{\tau_\Lambda}^0 d\tau_1 \int_{\tau_\Lambda}^{\tau_1} d\tau_2 \int_{-\infty}^{\tau_\Lambda} d\tau_3 \int_{-\infty}^{\tau_3} d\tau_4 H_{2}(\tau_1) H_{3}(\tau_2) \tilde H_{I} (\tau_3) \tilde H_{I} (\tau_4) | 0 \rangle \\
	& + 2 {\rm Re} \langle 0 | \phi\phi\phi \int_{\tau_\Lambda}^0 d\tau_1 \int_{\tau_\Lambda}^{\tau_1} d\tau_2 \int_{-\infty}^{\tau_\Lambda} d\tau_3 \int_{-\infty}^{\tau_3} d\tau_4 H_{3}(\tau_1) H_{2}(\tau_2) \tilde H_{I} (\tau_3) \tilde H_{I} (\tau_4) | 0 \rangle  ~.
	\end{align} 
	The first, second, fifth, sixth, ninth, tenth, elevent and twelfth line evaluates to the same form as the first line of \eqref{3ptbetabeta} plus its conjugate. The third, forth, seventh, eighth lines evaluate to the same form as second line of \eqref{3ptbetabeta} plus its conjugate
	
	\subsection{4pt} 
	In this subsection, we consider the in in formalism to compute the four-point correlation function. 
	\subsubsection{$\mathcal{O}(r)$ Order}
	\begin{figure}[h!] 
		\centering 
		\includegraphics[width=5cm]{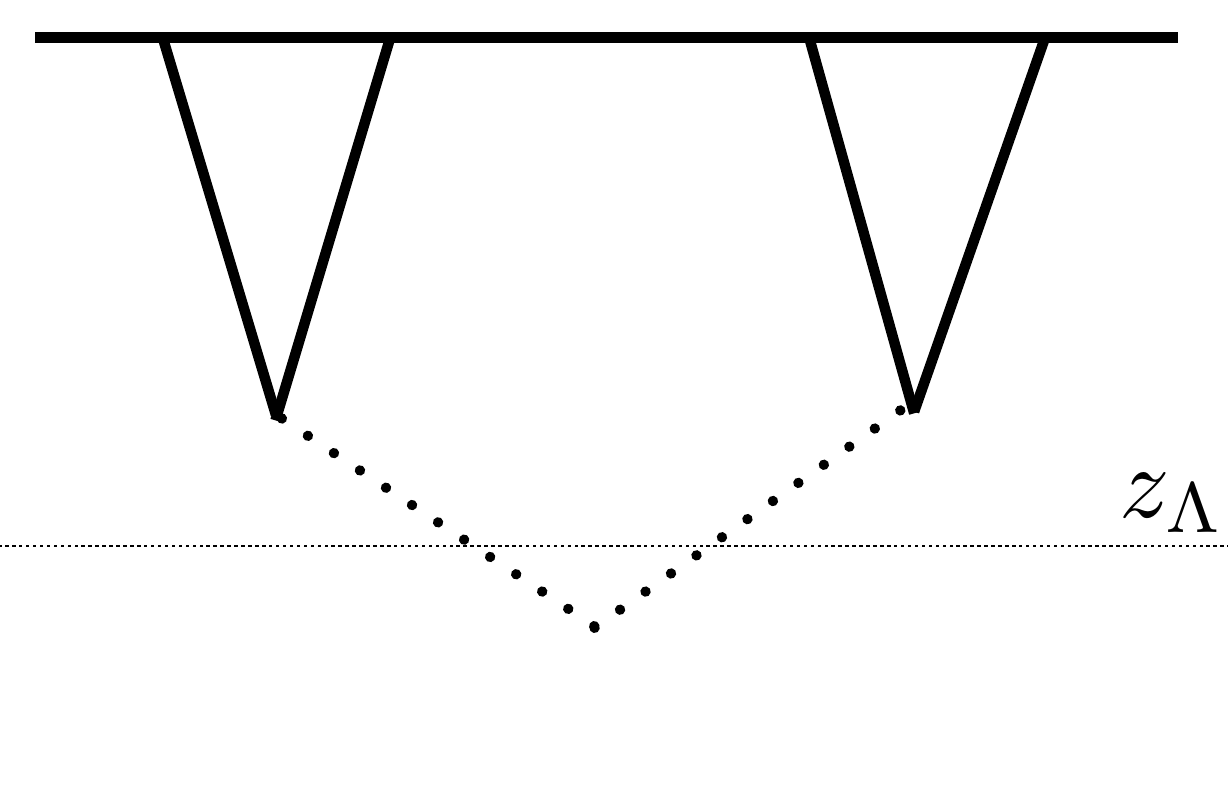}   
		\caption{Feynman diagram for the four-point correlation function at $r$ order. The upper horizontal line is the $\tau=0$ time slice, whereas the lower dashed horizontal line is the $k\tau=z_\Lambda$ cutoff time below which the effective mass square of the $\sigma$ field is $m^2+\Delta m^2$ and above which the effective mass of the $\sigma$ field is $m^2$. The other three solid lines represent the inflaton field $\phi$ and the dashed lines corresponds to the massive field $\sigma$. } \label{ininformalism4ptr1} 
	\end{figure}
	\begin{align} \nonumber
	\langle \phi\phi\phi\phi \rangle  = & -2 {\rm Im} \langle 0 | \phi\phi\phi\phi \int_{\tau_\Lambda}^0 d\tau_1 \int_{\tau_\Lambda}^{\tau_1} d\tau_2 \int_{-\infty}^{\tau_\Lambda} d\tau_3 H_{3}(\tau_1) H_{3}(\tau_2) \tilde H_{I} (\tau_3) | 0 \rangle  \\ \nonumber
	&  +  2 {\rm Im} \langle 0 | \int_{\tau_\Lambda}^0 d \tilde \tau_1 H_{3}(\tilde\tau_1) \phi\phi\phi\phi \int_{\tau_\Lambda}^0 d \tau_{1} \int_{-\infty}^{\tau_\Lambda} d\tau_2 H_{3}(\tau_1 ) \tilde H_{I} (\tau_2) | 0 \rangle \\
	&  +  2 {\rm Im} \langle 0 | \int_{-\infty}^{\tau_\Lambda} d \tilde \tau_1 \tilde H_{I} (\tilde\tau_1) \phi\phi\phi\phi \int_{\tau_\Lambda}^0 d \tau_{1} \int_{\tau_\Lambda}^{\tau_1} d\tau_2 H_{3}(\tau_1 ) H_{3}(\tau_2) | 0 \rangle  ~.
	\end{align} 
	The first line and the third line evaluate to the same form as \eqref{4ptalphabeta}. The second line is suppressed exponentially in the large $\mu$ limit.
	
	\subsubsection{$\mathcal{O}(r^2)$ Order}
	\begin{figure}[h!] 
		\centering 
		\includegraphics[width=5cm]{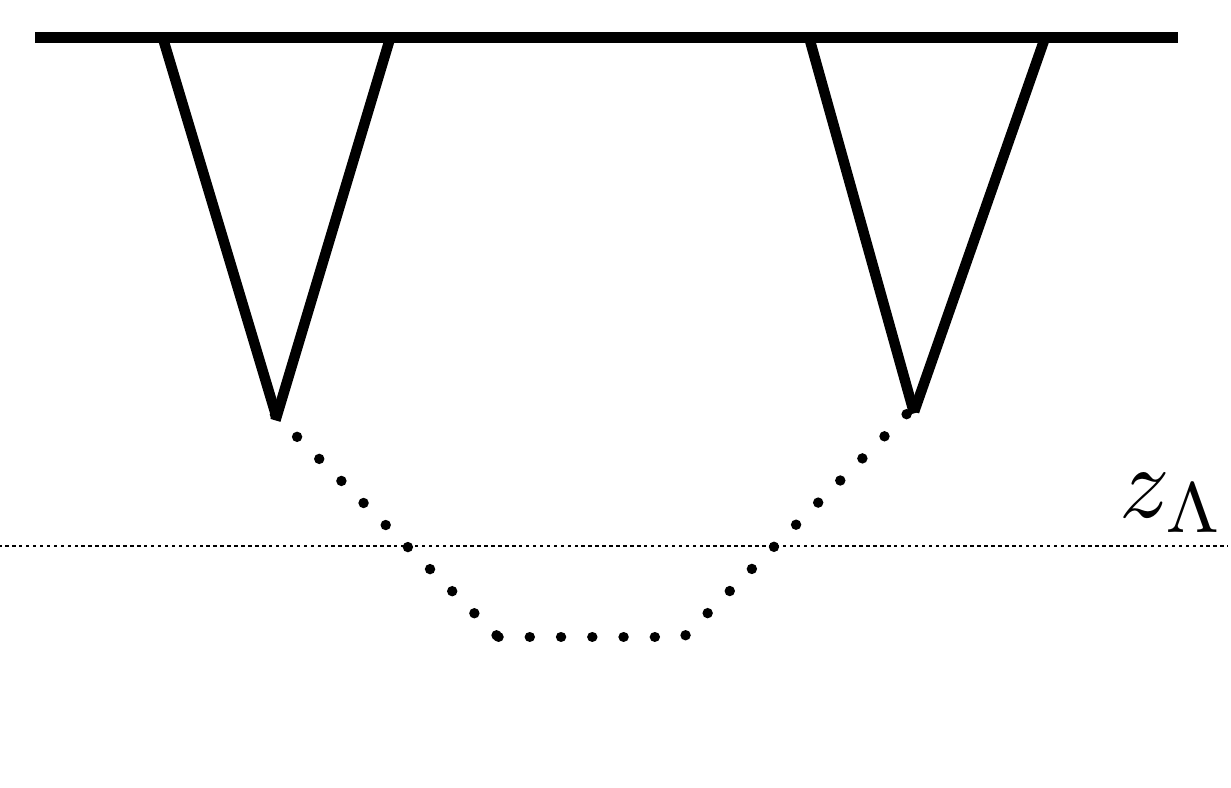}  
		\caption{Feynman diagram for the four-point correlation function at $r^2$ order. The upper horizontal line is the $\tau=0$ time slice, whereas the lower dashed horizontal line is the $k\tau=z_\Lambda$ cutoff time below which the effective mass square of the $\sigma$ field is $m^2+\Delta m^2$ and above which the effective mass of the $\sigma$ field is $m^2$. The other three solid lines represent the inflaton field $\phi$ and the dashed lines corresponds to the massive field $\sigma$. } \label{ininformalism4ptr2}  
	\end{figure}
	\begin{align} \nonumber
	\langle \phi\phi\phi\phi \rangle  & =   \langle 0 | \int_{\tau_\Lambda}^0 d\tilde \tau_1 \int_{\tau_\Lambda}^{\tilde \tau_1} d\tilde \tau_2 H_{3}(\tilde \tau_2) H_{3}(\tilde \tau_1)  \phi\phi\phi\phi \int_{-\infty}^{\tau_\Lambda} d\tau_1 \int_{-\infty}^{\tau_1} d\tau_2 \tilde H_{I}(\tau_1) \tilde H_{I}(\tau_2)  | 0 \rangle  \\ \nonumber
	& + \langle 0 | \int_{\tau_\Lambda}^0 d\tilde \tau_1 \int_{-\infty}^{\tau_\Lambda} d\tilde \tau_2 H_{I} (\tilde \tau_2) \tilde H_{3} (\tilde \tau_1)  \phi\phi\phi\phi \int_{\tau_\Lambda}^0 d\tau_1 \int_{-\infty}^{\tau_\Lambda} d\tau_2  H_{3}(\tau_1) \tilde H_{I}(\tau_2)  | 0 \rangle  \\ \nonumber
	& + \langle 0 | \int_{-\infty}^{\tau_\Lambda} d\tilde \tau_1 \int_{-\infty}^{\tilde \tau_1} d\tilde \tau_2 \tilde H_{I} (\tilde \tau_2) \tilde H_{I} (\tilde \tau_1)  \phi\phi\phi\phi \int_{\tau_\Lambda}^0 d\tau_1 \int_{\tau_\Lambda}^{\tau_1} d\tau_2  H_{3}(\tau_1) H_{3}(\tau_2)  | 0 \rangle  \\ \nonumber
	&  -  2 {\rm Re} \langle 0 | \int_{\tau_\Lambda}^0 d \tilde \tau_1 H_{3}(\tilde\tau_1) \phi\phi\phi\phi \int_{\tau_\Lambda}^0 d \tau_{1} \int_{-\infty}^{\tau_\Lambda} d\tau_2 \int_{-\infty}^{\tau_2} d\tau_3 H_{3}(\tau_1 ) \tilde H_{I} (\tau_2) \tilde H_{I} (\tau_3) | 0 \rangle \\ \nonumber
	&  -  2 {\rm Re} \langle 0 | \int_{-\infty}^{\tau_\Lambda} d \tilde \tau_1 \tilde H_{I}(\tilde\tau_1) \phi\phi\phi\phi \int_{\tau_\Lambda}^0 d \tau_{1} \int_{\tau_\Lambda}^{\tau_1} d\tau_2 \int_{-\infty}^{\tau_\Lambda} d\tau_3 H_{3}(\tau_1 ) H_{3}(\tau_2) \tilde H_{I}(\tau_3) | 0 \rangle \\
	& + 2 {\rm Re} \langle 0 | \phi\phi\phi\phi \int_{\tau_\Lambda}^0 d\tau_1 \int_{\tau_\Lambda}^{\tau_1} d\tau_2 \int_{-\infty}^{\tau_\Lambda} d\tau_3 \int_{-\infty}^{\tau_3} d\tau_4 H_{3}(\tau_1) H_{3}(\tau_2) \tilde H_{I} (\tau_3) \tilde H_{I} (\tau_4) | 0 \rangle   ~.
	\end{align} 
	The second line, and forth line evaluates to the same form as \eqref{4ptalphabeta}. The first line, third line, fifth and sixth line are suppressed in the large $\mu$ limit.

\end{document}